\documentclass[final,3p,times,authoryear]{elsarticle}

\usepackage{amssymb}

\usepackage{tabu}                      
\usepackage{booktabs}                  
\usepackage{lipsum}                    
\usepackage{mwe}                       
\usepackage{enumitem}
\usepackage{booktabs}
\usepackage{multirow}
\usepackage{rotating}
\usepackage{hyperref}
\usepackage{changepage}
\usepackage{float}
\usepackage{makecell} 
\usepackage{caption}
\usepackage{subcaption}
\usepackage{enumitem}
\usepackage{multirow}
\usepackage{rotating}
\usepackage{xcolor,colortbl}
\definecolor{Mint}{HTML}{D9F0E5} 
\definecolor{Brown}{HTML}{F6EEE9}
\definecolor{Butter}{HTML}{FFF6DA}
\definecolor{Lav}{HTML}{EEEAFB}

\usepackage{comment}
\usepackage{url}
\usepackage{natbib}

\journal{International Journal of Human-Computer Studies}

\begin{document}

\begin{frontmatter}



\title{What is the message? Perspectives on Visual Data Communication in Popular Science}

\author[label1,label2]{Regina Schuster}
\ead{regina.maria.veronika.schuster@univie.ac.at}

\author[label3]{Kathleen Gregory}
\ead{k.m.gregory@cwts.leidenuniv.nl}

\author[label1,label2]{Christian Knoll}
\ead{christian.knoll@univie.ac.at}

\author[label1,label4]{Torsten Möller}
\ead{torsten.moeller@univie.ac.at}

\author[label1,label5,label6]{Laura Koesten\corref{cor1}}
\ead{laura.koesten@mbzuai.ac.ae}

\affiliation[label1]{organization={Faculty of Computer Science, University of Vienna},
            addressline={Währinger Straße 29},
            postcode={1090},
            city={Vienna},
            country={Austria}}
\affiliation[label2]{organization={UniVie Doctoral School Computer Science DoCS, University of Vienna},
            addressline={Währinger Straße 29}, 
            postcode={1090},
            city={Vienna},
            country={Austria}}
\affiliation[label3]{organization={Centre for Science and Technology Studies (CWTS), Leiden University},
            addressline={Kolffpad 1},
            postcode={2333},
            city={Leiden},
            country={Netherlands}}
\affiliation[label4]{organization={Research Network Data Science, University of Vienna},
            addressline={Kolingasse 14-16},
            postcode={1090},
            city={Vienna},
            country={Austria}}
\affiliation[label5]{organization={Department of Human-Computer Interaction, Mohamed bin Zayed University of Artificial Intelligence},
            addressline={Masdar City},
            postcode={SE45 05},
            city={Abu Dhabi},
            country={United Arab Emirates}}
\affiliation[label6]{organization={AIT Austrian Institute of Technology},
            addressline={Giefinggasse 4},
            postcode={1210},
            city={Vienna},
            country={Austria}}

 \cortext[cor1]{Corresponding author}
 
\begin{abstract}
Data visualizations are widely used to communicate messages about urgent topics such as climate change and public health. However, we still know little about how these visualizations are produced and interpreted in popular science contexts. In this mixed-method study, we examine how data are visually communicated and understood in the popular science magazine Scientific American, focusing on the messages these visualizations convey.

To capture this complexity, we analyze data visualizations about climate change and pandemics in Scientific American over the past fifty years from three complementary perspectives: reader, chart, and producer. From the reader's perspective, we articulate takeaway messages and document sensemaking, interpreting visualizations first without and then with textual elements. From the chart perspective, we examine how visual features and text shape interpretation. From the producer's perspective, we draw on interviews with Scientific American staff to understand message planning and compare a sample of their intended messages with those we interpreted.

Using takeaway messages as our central analytic lens, we develop a message typology and show that messages vary systematically across dimensions such as granularity, articulation, and inference. A key finding is that text plays a pivotal role: approximately two-thirds of messages change when textual elements are added. While the interviews highlighted the central role of message planning in visualization production, intended and interpreted messages only partially aligned. Our findings underscore the importance of contextual clarity and audience-aware communication, and we derive recommendations for visualization designers and science communicators.
\end{abstract}




\end{frontmatter}

\section{Introduction}
\label{sec:introduction}

Data visualizations are used to communicate complex data to diverse audiences and are increasingly a part of people's daily lives \citep{engebretsen2020data,zhang2021mapping}. They are communicative tools that can shape public perception and understanding by influencing what people consider salient, trustworthy, or actionable, factors that, in turn, shape public discourse and societal decision-making. This is particularly relevant for climate change and public health, domains characterized by complex, uncertain data, high stakes, and dependence on broad public support to drive policy action. At the same time, both topics are often politicized and subject to misinformation \citep{solhart2020politicization}, underscoring the importance of clear, trustworthy communication. For such topics, popular science publications are a key interface between scientific knowledge and the public, and they increasingly use visual content \citep{duarte2022role,hiippala2020multimodal,amit-danhi2018digital}. Such content often migrates to news outlets and social media, amplifying reach and influence \citep{gregory2024data}.

Despite the prevalence of such visualizations, we know surprisingly little about how they are designed and interpreted. What messages do readers derive from such visualizations, and how do these messages vary? How do these messages depend on chart types and features, such as textual framing in the form of titles, captions, or surrounding article text? And how do producers' intentions relate to what readers ultimately take away? These questions span production, design, and interpretation---processes that are rarely studied together, yet are inseparable in practice.

Chart effectiveness has often been operationalized through measures that emphasize decoding, such as accurately reading a specific value from a chart, identifying trends in the data, or answering comprehension questions \citep{bolko2025numbers, burns2020how, north2006measuring1}. We argue that what data visualizations communicate to readers goes beyond what such measures capture, and that this communication cannot be reduced to accuracy alone.

Prior work has pointed toward `takeaway messages' as an analytic lens for communicative meaning. Messages capture what readers consider the main point, bridging producers' intent and readers' interpretation. What readers take away depends on producers' intentions and design choices, as well as on readers' prior knowledge, subjective experience, and interpretive judgments \citep{liu2010mental,karer2020conceptgraph}. Although deriving a takeaway message is integral to comprehension and sensemaking \citep{burns2020how,pinker1990theory,karer2020conceptgraph}, messages vary across readers, are rarely made explicit, and are difficult to measure \citep{knoll2025gulf,bearfield2024same,pinker1990theory}.

Messages are influenced by the entire chart, not just the data display. Visualizations combine visual elements with textual components such as titles, captions, labels, and annotations, often within the context of broader texts. A small but growing body of empirical work shows that textual components play a crucial role in how people engage with and remember visualizations, and that these processes vary across individuals \citep{stokes2026analysis,stokes2023role,borkin2016memorability}. Although the importance of text is increasingly recognized, relatively little work examines how textual and visual elements work together during sensemaking and shape what readers ultimately take away.

Understanding these processes requires examining three interconnected aspects: how readers articulate takeaway messages, how chart features and textual framing shape these messages, and how they relate to producers' intentions. To address this, we developed a mixed-method approach examining how readers make sense of visualizations, how visual and textual elements guide interpretation, and how producers plan intended messages. We use the written articulation of messages to surface and study these processes, informed by earlier work on data-centric sensemaking \citep{koesten2021talking,koesten2020everything}.

We focus on a popular science context, where visualizations play a key role in communicating complex scientific ideas to diverse audiences. To this end, we examine a corpus spanning more than 50 years of visual communication on climate change and pandemics in Scientific American, one of the longest-running English-language popular science magazines. To explore the messages these visualizations convey, we analyze 171 visualizations from 58 print and online articles from three complementary perspectives: reader, chart, and producer. From the reader's perspective, the research team articulates takeaway messages and documents sensemaking processes through field notes, first interpreting visualizations without accompanying textual elements (e.g., explanatory titles or captions), and subsequently with them. From the chart perspective, we characterize the corpus and examine how visual features and textual elements shape interpretation. Finally, we incorporate the producer perspective by drawing on interviews with 11 Scientific American staff to understand message planning practices and to compare a sample of intended messages with those we interpreted.

Across perspectives, our findings show that visualization messages are neither fixed nor solely determined by visual encoding but emerge through an interplay of design choices, textual framing, and audience sensemaking. From the reader's perspective, articulated messages varied in structure and granularity, often moving beyond description to interpretative or evaluative statements, and frequently emphasizing spatial, temporal, or change-related dimensions. From the chart perspective, visual encodings, complexity, and framing influenced interpretation, while textual elements played a decisive role: without accompanying text, the research team found chart interpretation difficult, and titles and annotations frequently shifted the level of detail and interpretative emphasis of the messages. From the producer's perspective, messages were typically planned early and refined collaboratively with audience needs and communication goals in mind, often operating at a higher level of abstraction and narrative framing. While many articulated messages aligned with intended takeaways, divergences reflected differences in emphasis and framing rather than simple misunderstanding. Together, these findings highlight visual data communication as a process in which meaning is shaped through the interaction of producers, visual and textual design, and reader interpretation.

This work advances understanding of visualization messages in public science communication and contributes to research in human–computer interaction, science communication, and data visualization. We present a multi-perspective analysis of how messages are conveyed and interpreted in public-facing visualizations. The principal contributions are:
\begin{itemize}[itemsep=0.1pt]
    \item a message typology describing how takeaway messages are articulated;
    \item a characterization of a longitudinal corpus of visual data communication for public audiences;
    \item insights into how chart features, textual elements, and producer intentions shape visualization sensemaking and interpretation;
    \item an exploratory mixed-method approach to examining visualization messages across production and reception;
    \item design recommendations for data visualization and science communication aimed at public audiences.
\end{itemize}

\section{Background}
\label{sec:background}
We situate our work within the following research areas. We first discuss data visualization in public science communication (Section~\ref{subsec:background-datavis}). We then review sensemaking research on data visualizations (Section~\ref{subsec:background-sensemaking}) with a closer inspection of i) takeaway messages as an analytic lens for studying visualization interpretation (Section~\ref{subsec:background-sensemaking-message}) and ii) the role of textual elements in shaping interpretation (Section~\ref{subsec:background-sensemaking-text}).

\subsection{Data visualization in public science communication}
\label{subsec:background-datavis}

Public science communication has been integral to scientific research since its earliest days \citep{shapin1990science} and helps bridge the gap between scientific communities and the broader public. Research on popular science outlets has examined a range of content types, including images \citep{born2020visual,heekeren2021popular, gruber2012persuasive, hornmoen2010making}, text and writing styles such as storytelling \citep{molek-kozakowska2024making}, metaphors \citep{pramling2007scientific}, infographics \citep{lee2022perceptions,teixeira2018digital}, and cinematic or 3D data visualization \citep{jensen2024picture, jensen2022new}. Strategies such as humanization and aesthetization are commonly used to make scientific content accessible and engaging \citep{molek-kozakowska2019making, molek-kozakowska2024making}. Public science communication faces particular challenges in domains characterized by complex, uncertain data and contested public trust, such as climate change and public health. Both topics are subject to politicization and misinformation \citep{ecker2022psychological,solhart2020politicization}, making clear and trustworthy visual communication especially important. Visualizations on these topics frequently migrate beyond their original publication context to news outlets and social media \citep{shi2025double, gregory2024data}, where they may be encountered without the surrounding article context that supports interpretation and face distinct design requirements \citep{ramos2025science,shang2024exploration}.

An overview of the European science communication landscape \citep{rachaeldavies2021landscape} suggests that traditional forms of science communication are evolving, particularly in recent years, to incorporate new media and channels, such as social media \citep{ramos2025science}. This shift has also led to an increase in the types of visuals and to a wider use of data visualizations across different contexts, including in popular science magazines \citep{hiippala2020multimodal,amit-danhi2018digital}. Despite their acknowledged importance in science communication \citep{duarte2022role, engebretsen2020whatworkcan} and their potential to reduce barriers posed by text-heavy content \citep{perra2021seeing}, data visualizations in popular science have received less research attention than other content types. One notable exception is \citet{krause2017framework}, which examines how Nature adapts its figures and data visualizations to different audiences and content types given its mix of peer-reviewed and explanatory content.

Scientific American, the oldest continuously published magazine in the English-speaking world, is situated between scientific journals and mass media in the science communication landscape. Alongside other popular science magazines (e.g., New Scientist or the German-language GEO) and more general outlets that regularly publish on climate change and COVID-19 (e.g., The Atlantic or The New Yorker), Scientific American offers dedicated science-related coverage to educated and interested audiences. Available both in print and online formats, it reaches millions of people globally each month \footnote{\url{https://www.scientificamerican.com/page/about-scientific-american} (retrieved in February 2026)}. While the language used by Scientific American has been the subject of prior research \citep{fjohnston2018vaunting,ye2021abstracts,kirk2021visualization}, its data visualizations remain largely unstudied, leaving open questions about how complex scientific topics are communicated visually to public audiences. This paper is part of a broader research project on data visualizations in Scientific American. In related work, we examine the production context of visualizations from a socio-technical perspective, including the practices and decisions through which they are created \citep{gregory2024data}. The present paper instead examines how data visualizations, including their design choices, communicate meaning across production and interpretation, using articulated takeaway messages as an analytic lens.

\paragraph{Summary} Despite the acknowledged importance of data visualizations for communicating complex, contested topics such as climate change and public health, they have received comparatively little research attention in popular science contexts. Existing work has focused primarily on other content types, and the few studies that examine visual communication practices mostly focus on specialist rather than popular science outlets. To our knowledge, no study has examined how data visualizations in popular science communicate meaning across production and interpretation, or how this has evolved over time.

\subsection{Sensemaking and communicative meaning in data visualization}
\label{subsec:background-sensemaking}

Data visualizations are not neutral representations of data but communicative artifacts shaped by rhetorical choices, design decisions, and contextual framing \citep{correll2019ethical,hullman2011visualization,kostelnick2008visual}. Much of the research on visualization interaction has focused on perceptual studies investigating visual encodings of specific data types to tailor chart design to our visual system (see \cite{franconeri2021science} for a review). Effectiveness has largely been measured through accuracy-based tasks such as value retrieval, trend identification, and comparison \citep{ridley20208, burns2020how, lee2017vlat}. With a few notable exceptions \citep{lee2016how, pohl2017sense-making, baker2009using, burns2022communicative, wang2019comparing}, people's ability to make sense of visualizations and what they take away as communicative meaning remains less explored. 

Making sense of data visualizations is a complex challenge requiring multiple skills and literacies simultaneously. The general concept of sensemaking has been studied across different disciplines, such as psychology \citep{klein2006making}, human-computer interaction \citep{pirolli2005sensemaking, mrussell1993cost}, and education \citep{urquhart2024sense-making}, and involves linking and constructing meaning from different pieces of information into a single conceptual representation \citep{mrussell2003learning, hearst2009search}. Data visualizations add further challenges to this broader tradition: they are embedded in the epistemology of data science \citep{neff2017critique} and require readers to decode graphical and textual elements while conceptually placing data within contexts of creation, disciplinary norms, and the world as a whole \citep{koesten2021talking}. This process draws on multiple types of literacy, i.e., visual, data, and textual literacy, and involves a complex interplay of bottom-up and top-down cognitive processing \citep{karer2020conceptgraph}, encompassing encoding, decoding, and reasoning activities \citep{pirolli2005sensemaking, baker2009using}.

Because sensemaking draws on prior knowledge, experience, and interpretive strategies, the same visualization can be interpreted in substantially different ways depending on the reader \citep{lee2016how, boy2014principled}. Readers do not simply decode what is visually present but actively construct meaning, which is inherently influenced by human factors such as visual attention, perception, and judgment \citep{alhadad2018visualizing}. Understanding what readers actually take away from visualizations, and how this relates to what producers intend to communicate, therefore requires looking beyond accuracy-based measures toward the communicative meaning readers construct.

\subsubsection{Message articulation as an analytic lens for studying sensemaking}
\label{subsec:background-sensemaking-message}

Messages can be defined as individual summaries of the key content of a chart, as interpreted by the reader or intended by the producer, and encompass not just data patterns but also social, emotional, and evaluative dimensions of meaning. While interest in messages as an analytic lens has grown considerably in recent years \citep{stokes2025write,shi2025double, bolko2025numbers, knoll2025gulf, schuster2024being, burns2022invisible, fernandeznieto2022learning, ajani2021declutter, zhang2021mapping, burns2020how, ballantyne2018exploring, bateman2010useful}, empirical work on how individuals construct overall meaning from visualizations remains limited, and little is known about how this process unfolds in its full complexity, particularly across different contexts and audiences. Most existing research has focused on performance-based tasks such as value retrieval and comparison, which do not capture the fuller communicative meaning readers construct \citep{bolko2025numbers}. Examining articulated messages, therefore, offers a window into sensemaking that goes beyond what such measures can capture \citep{burns2020how}.

Research on visualization processing suggests that the messages readers construct from visualizations emerge from an interaction between the chart's design properties and their prior knowledge, beliefs, and biases \citep{schloss2025perceptual}. On the reader side, messages are co-constructed rather than fixed: what readers take away shifts depending on their interpretive strategies and the context in which they encounter the visualization \citep{adar2021communicative}. Messages can vary substantially across individuals, with expertise playing a particularly important role \citep{schuster2024being}, and can extend well beyond the data explicitly encoded, including social inferences about the artifact's provenance and trustworthiness \citep{fox2025quantifying, morgenstern2025visualization}. On the design side, choices such as the ordering, spacing, and coloring of marks directly shape which messages a visualization affords, such that the same data can communicate substantially different messages depending on its arrangement \citep{fygenson2023arrangement, xiong2021visual, franconeri2021science}. Yet existing work has largely examined these factors in isolation and using simplified, controlled stimuli; how design properties and reader factors interact to shape messages articulated from complex, real-world visualizations, and in particular what role textual elements play in that process, remains poorly understood.

The importance of articulating a clear message is recognized in data visualization practice: studies of visualization producers suggest that communicating a specific message or goal is a central consideration in chart design \citep{schuster2026practitioners, zhou2024epigraphics, parsons2021understanding}. Practitioners pursue a range of communicative goals through their messages, including capturing attention, supporting comprehension, building trust, and evoking emotion \citep{schuster2026practitioners}. A particular strand of work has focused on persuasive messages, i.e., the ways in which visualization design choices such as highlights, annotations, and framing can shift the message a chart conveys to support particular perspectives or attitudes \citep{markant2023when,markant2022can,pandey2014persuasive}. Recent work in visualization authoring has begun to acknowledge this message focus through message-first workflows that use intended messages to guide and generate visual components \citep{zhou2024epigraphics}. This message-first orientation is also central to narrative visualization and data storytelling \citep{shao2024data}, where planning the intended message before designing visual representations is recognized as a defining feature of the narrative-oriented process \citep{zhang2022visual}. Despite the breadth of research on visualization practice \citep{parsons2021understanding,schuster2026practitioners}, how practitioners conceptualize, plan, and work with messages in their design process has received little explicit attention.

Nonetheless, even when producers design with a clear message in mind, intended and interpreted messages do not necessarily align: \citet{knoll2025gulf} compared producer-encoded messages with reader interpretations and found what they call ``a gulf of interpretation'' between the two, shaped by data density, unfamiliar conventions, and differences in expertise. This misalignment is not simply a failure of design; it reflects the fundamentally co-constructed nature of sensemaking, in which reader and design factors interact in ways that are difficult to anticipate. Surfacing messages can therefore serve as a way of creating common ground between readers and producers \citep{clark1989contributing, stalnaker2002common}, making explicit what each group takes to be the key content of a visualization and providing an empirical basis for understanding where and why intended and interpreted messages diverge.

Studying messages empirically, and thus making this common ground accessible, relies on summarization as both a concept and a method: articulating a message from a visualization is inherently an act of summarization, distilling the key content of a chart into a concise statement. Summarization has a well-established role in meaning-making research: it has been studied as a cognitive process in both text comprehension \citep{hidi1986producing, kintsch1978model} and data understanding, where both textual and verbal summaries have been used to assess how people make sense of data \citep{koesten2021talking, koesten2020everything}. In the context of visualization specifically, free-response tasks have been identified as an effective method for capturing the messages readers tend to extract from visualizations, precisely because they provide rich, unconstrained information about reader interpretations \citep{stokes2025write}. However, meaningful abstractive summaries are inherently subjective: there are no objective criteria for what counts as the most important thing to include \citep{gambhir2017recent}. We address this challenge by treating messages as analytic outputs rather than objective ground truth, focusing on recurring patterns across articulated messages to understand how communicative meaning is constructed and varies across readers and conditions.

\subsubsection{Role of textual elements in sensemaking of data visualizations}
\label{subsec:background-sensemaking-text}

Textual elements in and around data visualizations, such as titles, captions, annotations, and labels, are integral components of how charts communicate meaning to readers \citep{segel2010narrative, hullman2011visualization}. They serve a wide range of functions from identifying data mappings to narrative framing \citep{stokes2026analysis}. Despite their ubiquity, the relationship between textual and visual elements has received comparatively little systematic attention in visualization research, although recent work is beginning to address this gap \citep{stokes2026analysis, stokes2025combining}. Text and visuals interact in complex, context-dependent ways: readers can struggle to integrate the two when both are present \citep{ottley2019curious}, the spatial arrangement and degree of linking between text and visuals affects comprehension and engagement \citep{zhi2019linking}, and whether text or visuals plays the more important role in shaping understanding varies across readers and contexts \citep{stokes2023role, hearst2019would, mckenna2017visual}. Yet how these complex text-visual interactions shape the messages readers construct from real-world visualizations, particularly in public science communication contexts, has not been systematically investigated.

Textual elements shape what readers take away: the semantic content and placement of text directly influence the types of messages readers construct from visualizations, and framing effects introduced through titles and captions can substantially alter the perceived tone and meaning of a chart \citep{stokes2023role}. Titles in particular have an outsized influence: readers' recall of a visualization's message tends to align more with the title than with the visual content itself, even when the two contradict each other \citep{kong2018frames, kong2019trust}. Captions shape which messages are emphasized, with visual prominence and textual emphasis interacting in complex ways \citep{kim2021towards, kim2023e}. Annotations similarly allow visualization creators to deliberately direct reader attention toward particular messages \citep{ren2017chartaccent, bearfield2024same, rahman2025survey}. Beyond immediate interpretation, text has also been shown to be a key driver of visualization memorability \citep{borkin2016memorability, he2024visual}.

Beyond their influence on reader takeaways, textual elements are actively preferred by readers: charts with more annotations have been shown to be favored over minimally annotated ones, with readers valuing being informed even at the cost of simplicity \citep{stokes2022striking, stokes2022why}. Yet preferences do not straightforwardly translate into better comprehension; informative titles reduce mental effort compared to generic ones but do not necessarily improve accuracy or credibility \citep{wanzer2021role}. What makes textual elements meaningful and effective from a reader's perspective, therefore, remains a question that existing work has only begun to address.

\paragraph{Summary} Despite a growing body of work on visualization sensemaking, most research has focused on perceptual accuracy and low-level comprehension tasks, leaving the communicative meaning readers construct from visualizations largely underexplored. Takeaway messages offer a promising analytic lens for studying this meaning-making process, but existing work has not systematically examined the types of messages readers articulate, how they are shaped by textual elements, or how they relate to producers' intended communication. While the role of text in shaping interpretation is increasingly recognized, its effects on overall message takeaway, particularly in complex, real-world public science communication contexts, have received little systematic attention. Finally, while the importance of a clear intended message is widely acknowledged in visualization practice, how producers conceptualize and work with messages remains understudied. This study addresses these gaps by examining messages across reader, chart, and producer perspectives in a corpus of real-world public science visualizations spanning fifty years of climate change and pandemic coverage, with an emphasis on climate change.
\section{Methodology}
\label{sec:methodology}

We designed a multi-step exploratory methodology to investigate the following research questions regarding visualizations on climate change and pandemics in Scientific American:
\begin{itemize}
    \item \colorbox{Mint}{\textbf{Reader perspective:}} How do readers form and articulate `messages' when engaging with data visualizations?
    \item \colorbox{Brown}{\textbf{Chart perspective:}} Which visualization types are used, which design characteristics and textual elements are present in the charts, and how do these relate to readers' sensemaking and the messages they articulate?
    \item \colorbox{Lav}{\textbf{Producer perspective:}} How do visualization producers conceptualize and plan messages during visualization design, and how do these intended messages relate to those articulated by readers?
\end{itemize}

\begin{figure*}[htb]
    \centering
    \includegraphics[width=\textwidth]{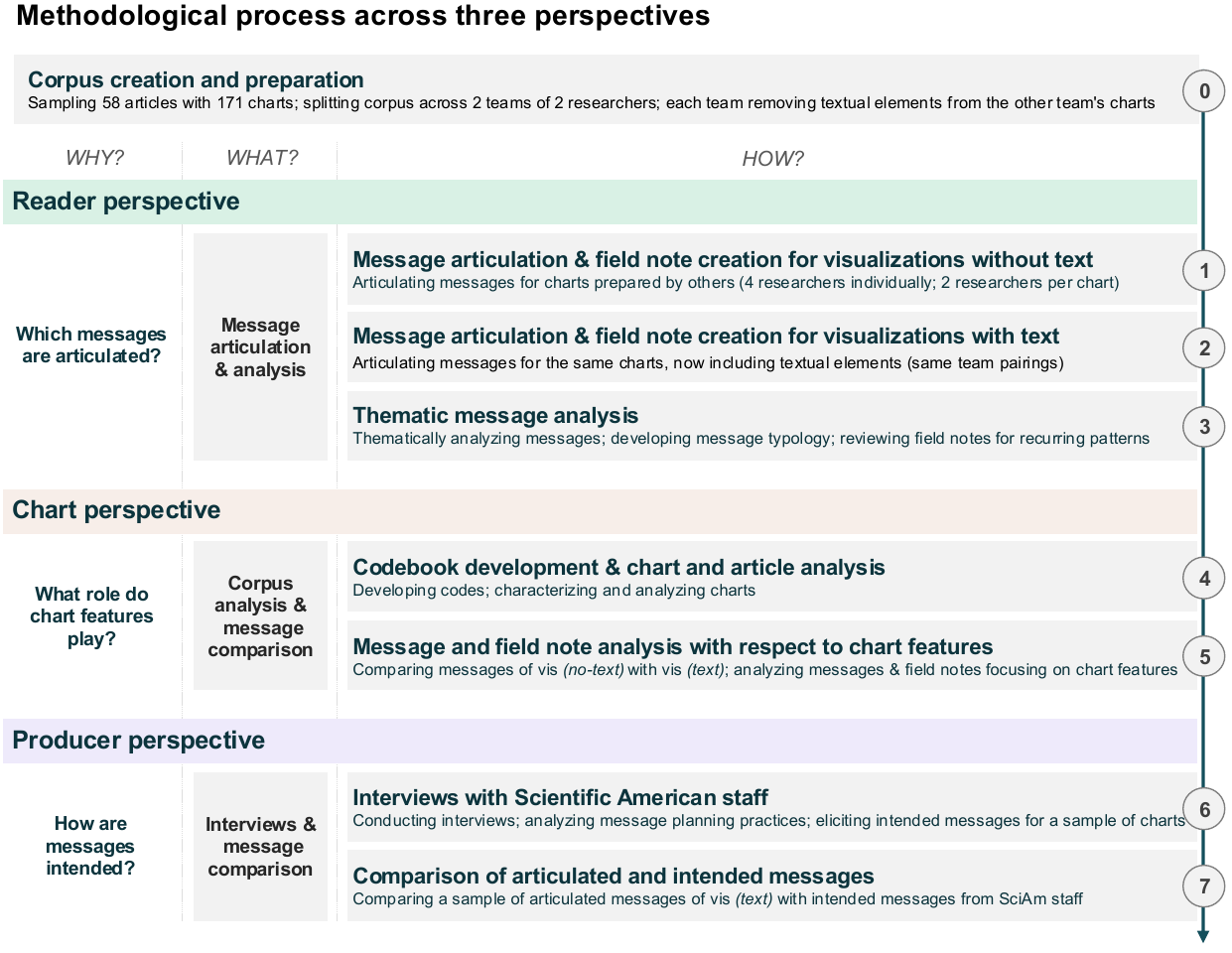}
    \caption{Overview of the methodological process across three perspectives. Following a shared preparation step (0) in which each team removed textual elements from the other team's charts, the study proceeded through seven steps spanning the reader (1-3), chart (4-5), and producer perspectives (6-7).}
    \label{fig:methods}
\end{figure*}

To address these questions, a team of four authors analyzed 171 visualizations from 58 articles published in Scientific American over a 50-year period. We first interpreted these charts without textual elements, then with them, to identify and articulate the main messages they communicated to us, documenting our sensemaking throughout the interpretation process in field notes \textit{(reader perspective)}. Next, we coded the features of these visualizations to characterize the chart corpus and analyze the articulated messages in relation to chart design \textit{(chart perspective)}. Finally, we interviewed Scientific American staff to better understand the creation of intended messages and their relationship to visualization design \textit{(producer perspective)}. \autoref{fig:methods} provides an overview of our study design from those three perspectives: reader, chart, and producer. This study is part of a project that has undergone ethical review in accordance with our institution's guidelines and has been determined to be low risk.

\subsection{Corpus creation and preparation}
\label{subsec:methodology-preparation}
Before the main methodological steps, we defined the article sample, established the visualization corpus, and prepared the visualizations extracted from Scientific American articles.

\paragraph{Sampling strategy} We partnered with Scientific American to obtain 100 print and online articles containing data visualizations. The articles were published between 1918 and 2022 and covered climate change and pandemics, including COVID-19, influenza, and other infectious diseases. To capture both current practices and longitudinal change, we prioritized recent articles while also sampling across earlier decades. We therefore included all articles published between 2017 and 2022 (n=43) and, in addition, randomly sampled articles from 1970 to 2016 (n=15), stratified into 5-year groups. Only three articles were published before 1970 and were excluded from the analysis.\footnote{The three excluded articles were from disparate years: one was published in 1918 and one in 1959; the third article was missing a visible date.} This resulted in a final corpus of 58 articles.

\paragraph{Corpus composition} Most articles covered climate change (n=49); nine pertained to pandemics, of which eight focused specifically on COVID-19. This subject distribution reflects the recency of the COVID-19 pandemic and the long-standing coverage of climate topics in Scientific American. In terms of format, 59\% of articles were originally published in print, including longer in-depth feature pieces (n=27) and one-page graphic-centered articles (n=8); the remaining 23 articles were shorter, online news-oriented pieces. A single article could contain multiple visualizations; for example, one COVID-19 article contained 22 individual charts, yielding a substantially larger total number of visualizations across the 58 articles. The supplementary material provides a full breakdown of the article corpus by subject, type, and publication year (S1) as well as an overview of all article references (S3).

\begin{figure*}[htpb]
    \centering
    \includegraphics[width=\textwidth]{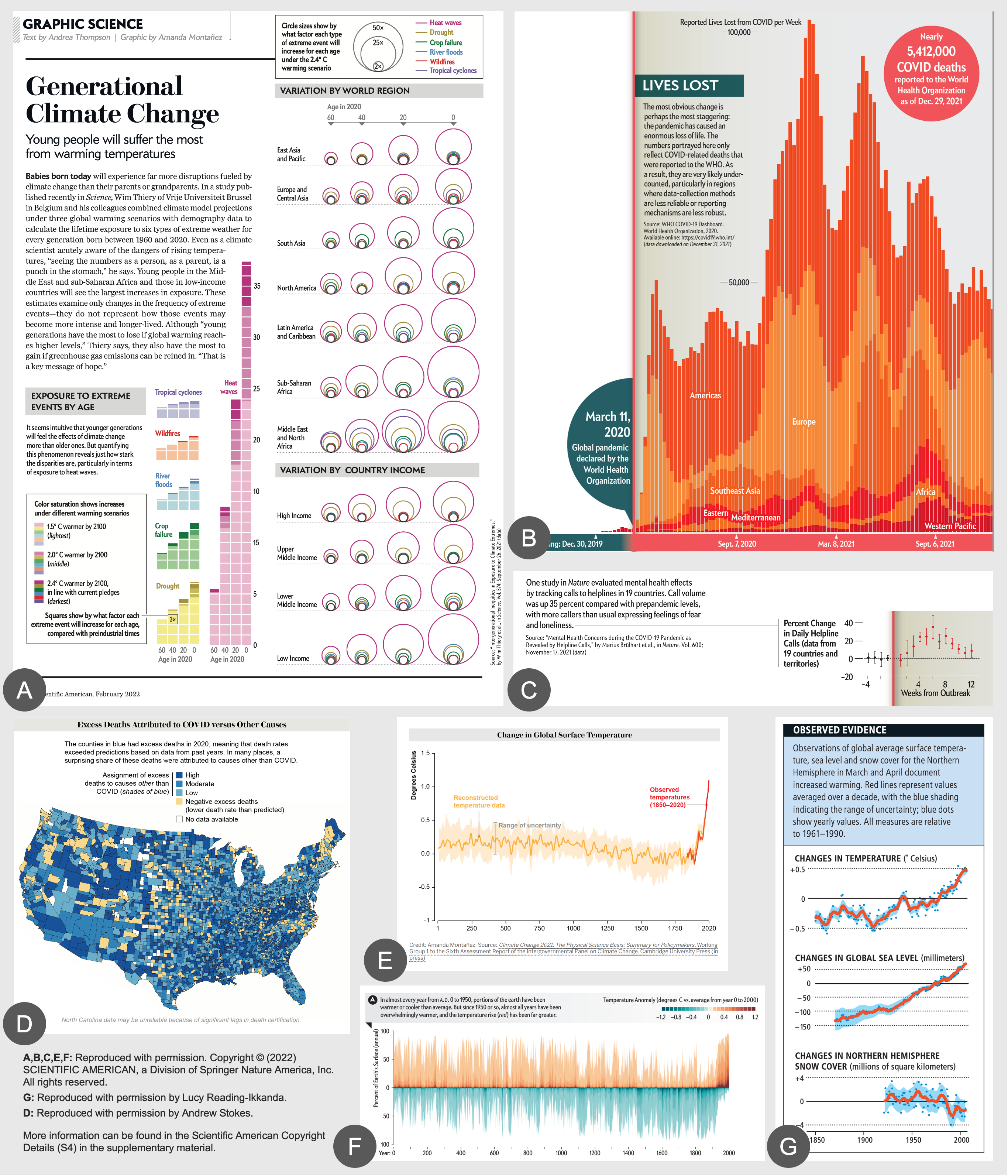}
    \caption{Example charts. Article references can be found in the Scientific American Article Bibliography (S3) in the supplementary material: A [56], B [33], C [33], D [27], E [58], F [16], G [9]. \textit{(A,B,C,E,F: Reproduced with permission. Copyright © (2022) SCIENTIFIC AMERICAN, a Division of Springer Nature America, Inc. All rights reserved.; D: Reproduced with permission by Andrew Stokes.; G: Reproduced with permission by Lucy Reading-Ikkanda. More information can be found in the Scientific American Copyright Details (S4) in the supplementary material.)}}
    \label{fig:examplecharts}
\end{figure*}

\paragraph{Visualization preparation} The 58 articles were divided between two teams of two researchers each. Each team manually extracted data visualizations from their assigned articles, yielding a total of 171. We defined data visualizations as \textit{any visual representation of data, excluding photographic images, tables, and explanatory diagrams}. Each team then stripped their assigned visualizations of textual elements, producing two versions of each: `vis \textit{(text)}', retaining all textual elements as displayed in the article, and `vis \textit{(no-text)}', with textual elements removed. Removed textual elements included captions, explanatory or narrative text, and titles; labels and legends were retained. An example of a chart with and without textual elements is shown in \autoref{fig:casestudy}. After preparation, teams switched corpora, ensuring that researchers articulated messages for charts they had not personally prepared. Each visualization was then independently interpreted by two researchers, first without and then with textual elements, yielding four articulated messages per visualization: two from the no-text condition and two from the text condition. Example charts illustrating the corpus are shown in \autoref{fig:examplecharts}.

\subsection{Reader perspective: Message articulation and analysis}
\label{subsec:methodology-reader}
To examine how messages can be articulated from data visualizations, we adopted a reader perspective focused on how meaning is constructed during chart interpretation. We performed a structured task in which we articulated a main message for each visualization. We define messages as a chart's main takeaway or primary communicative intent, with multiple takeaways possible; these messages are treated as analytic outputs rather than as proxies for real reader interpretations. In applying this approach, we acted in a dual role as readers engaging with the charts and as researchers documenting and reflecting on this process through field notes, drawing on principles of analytic autoethnography \citep{anderson2006analytic}.

\paragraph{Message articulation and field note creation} First, each researcher wrote a perceived main message for each `text-free' visualization, here referred to as the message of vis \textit{(no-text)}. Next, we articulated messages for vis \textit{(text)}, based on the original visualizations as they appeared in the article, including all textual elements. This process was done independently by two researchers per chart, resulting in 342 messages of vis \textit{(no-text)} and 342 messages of vis \textit{(text)}. We also created detailed field notes \citep{emerson2011writing} documenting our sensemaking and message articulation processes, including questions for the chart producers, observations and reflections on the messages and framing, and notes on our emotional engagement with the visualizations. 

We adopted this approach for several reasons. First, articulating messages ourselves allowed us to surface sensemaking processes in ways that are difficult to capture through external participants, who may not fully articulate their interpretive steps in the moment. Second, the scale of the corpus (171 visualizations interpreted in two conditions) would have been logistically difficult to cover systematically with external participants, and our sustained engagement with the full corpus enabled the kind of comparative perspective that underpins our findings. Third, message articulation and typology development could be refined iteratively, with insights from later visualizations informing the interpretation of earlier ones. We acknowledge that our shared disciplinary background introduces a bias toward expert interpretation; however, interviews with Scientific American staff confirmed that the research team is representative of a segment of their target audience with high data visualization expertise. We therefore argue that this expertise biases our results toward sensemaking success, and that a more general audience may show a stronger discrepancy between articulated and intended messages.

\paragraph{Thematic message analysis} Messages (for both vis \textit{(no-text)} and vis \textit{(text)}) and the accompanying field notes were thematically analyzed \citep{braun2019reflecting}. Individual articulated messages served as the primary unit of analysis, while field notes were used to contextualize and support theme development. Themes for message types and content were primarily inductively developed through open coding, but were also informed by literature on data summarization \citep{koesten2021talking}. This analysis led to the development of a message typology, as described in the findings section.

\subsection{Chart perspective: Corpus analysis and message comparison}
\label{subsec:methodology-chart}
To complement the reader perspective, we analyzed the visualizations from a chart perspective, aiming to characterize chart features, i.e., visualization types, design characteristics, and textual elements, across the corpus to identify patterns in visual data communication and examine how chart features relate to articulated messages.

\paragraph{Codebook development and chart and article analysis} We developed a codebook drawing on existing typologies for chart classification \citep{chen2022not}, visualization tasks \citep{jilaniquadri2021survey,brehmer2013multi-level}, and types of data encodings \citep{munzner2014visualization}. We organized codes into three categories: descriptive codes capturing chart types and types of text elements; semantic codes describing chart characteristics informed by literature (as exemplified in Table S2.T1 in the supplementary material); and subjective codes capturing chart framing \textit{(negative/neutral/positive/other)} and perceived coding effort \textit{(easy/neutral/hard)}. Over a 2.5-month period, the codebook underwent three rounds of iteration with the entire research team, including a senior data visualization expert (who was not part of the coding team). We thereby applied a combination of deductive coding from existing typologies and inductive development of new codes \citep{braun2019reflecting}. The codebook is organized into four groups based on analysis level: articles, charts, data, and text. All codes and sub-codes are described in the supplementary material (S2.T2).

The two teams of two researchers each (hereafter referred to as coders, C1--C4) coded visualizations using Atlas.ti, each coding the visualizations for which they had articulated messages in previous steps, resulting in each visualization being coded independently by two coders. Rather than reading entire articles, we coded shorter summarizing text blocks such as abstracts and call-out boxes, as our focus was on the visualizations themselves and the textual elements directly associated with them, rather than the full article content. After independent coding, each team held six one-hour consensus meetings to compare chart characterization codes such as chart type, dimensionality, and expressiveness. Differences were discussed within the full research team. Given the complexity of many characterization codes, we used consensus coding rather than quantitative agreement measures, treating disagreements as prompts for deeper discussion and collaborative theme development \citep{mcdonald2019reliability}, and reached consensus for all chart characterization codes. Codes related to framing and coding effort were not required to reach consensus, as they were designed to capture individual experiences and opinions.

\paragraph{Message and field note analysis with respect to chart features} Finally, from a chart perspective, we investigated how sensemaking and message articulation relate to chart features, i.e., chart types, design characteristics, and textual elements. We analyzed how particular chart types and characteristics influenced our sensemaking process. For each visualization, we also compared messages articulated for vis \textit{(no-text)} and vis \textit{(text)} to determine whether adding textual elements beyond those required to interpret the chart structure, such as explanatory annotations or titles rather than axis labels or legends, changed the articulated message. Additionally, two senior researchers closely reviewed the field notes and inductively developed themes capturing recurring patterns in sensemaking and reflection.

\subsection{Producer perspective: Interviews and message comparison}
\label{subsec:methodology-producer}
To complement the reader and chart perspectives, we drew on chart producers' perspectives through interviews, comparing intended messages with our articulated messages for a sample of visualizations.

\paragraph{Interviews with Scientific American staff} We conducted 11 one-hour semi-structured interviews with current and former Scientific American staff members; further details about the interview study are described in \cite{gregory2024data}. To protect participants' anonymity, we describe them at an aggregate level: participants (P) worked as text editors, graphics editors, visualization designers, and in leadership positions. Most were highly experienced: seven had over 15 years in their roles, three had 6-15 years, and one had 1 year. We used snowball sampling, recruiting all Scientific American staff involved in visual data communication who were suggested to us and were willing to participate.

Before the interview, participants selected an article with visualizations they had worked on or contributed to about climate change or COVID-19. We chose this approach to allow participants to speak from direct experience rather than reconstructing unfamiliar work from memory. During the interview, we asked participants to describe the process of producing the article and its visualizations and to reflect on envisioned messages and intended audience engagement. Each interview protocol was tailored to participants' expertise and professional role; the basic protocol is available in the supplementary material (S5). We analyzed coded transcripts using reflexive thematic analysis \citep{braun2019reflecting}, combining deductive and inductive coding.

\paragraph{Comparison of articulated and intended messages} As a final step, we asked participants who had worked on visualizations in our corpus to provide their intended messages for a sample of charts. In total, Scientific American staff provided intended messages for 18 visualizations. We compared these with our articulated messages through a qualitative thematic comparison, examining alignment and divergence in content, emphasis, framing, and level of abstraction. In this paper, we focus on interview findings related to the planning of visualization messages and the comparison between intended and articulated messages; a detailed analysis of the interviews is available in \cite{gregory2024data}. 
\section{Findings}
\label{sec:findings}
The findings are structured along the three perspectives outlined in Section~\ref{sec:methodology}, corresponding to our research questions, each drawing on a distinct analytical source: articulated messages and accompanying sensemaking field notes (reader perspective, Section~\ref{subsec:findings-reader}); the article corpus, including 171 visualizations (chart perspective, Section~\ref{subsec:findings-chart}); and interviews with visualization producers (producer perspective, Section~\ref{subsec:findings-producer}). Together, these analyses address how visualization takeaway messages are articulated, how chart features, particularly textual elements, shape these messages, and how reader sensemaking relates to producers' intentions and message planning practices. Throughout the findings, we refer to the four research team members involved in message articulation, field note creation, and coding as coders (C1--C4). While the findings reflect consistent trends across the corpus, individual coders at times responded differently to the same visualization, reflecting, for example, differences in prior knowledge, attention, and interpretive strategies. As our interest lies in understanding message variation as a broader phenomenon rather than in systematic cross-coder comparison, we discuss such individual differences where they are particularly informative.

To illustrate the seven analytical steps (\autoref{fig:methods}) with concrete examples, we selected four charts as illustrative case studies, focusing in particular on the message analysis: comparisons between messages articulated for vis \textit{(no-text)} and vis \textit{(text)}, and between messages articulated for vis \textit{(text)} and the intended messages shared by Scientific American staff. \autoref{fig:casestudy} illustrates Case Study I; all four case studies are included in the supplementary material (S7).

\begin{figure*}[htb]
    \centering
    \includegraphics[scale=0.65]{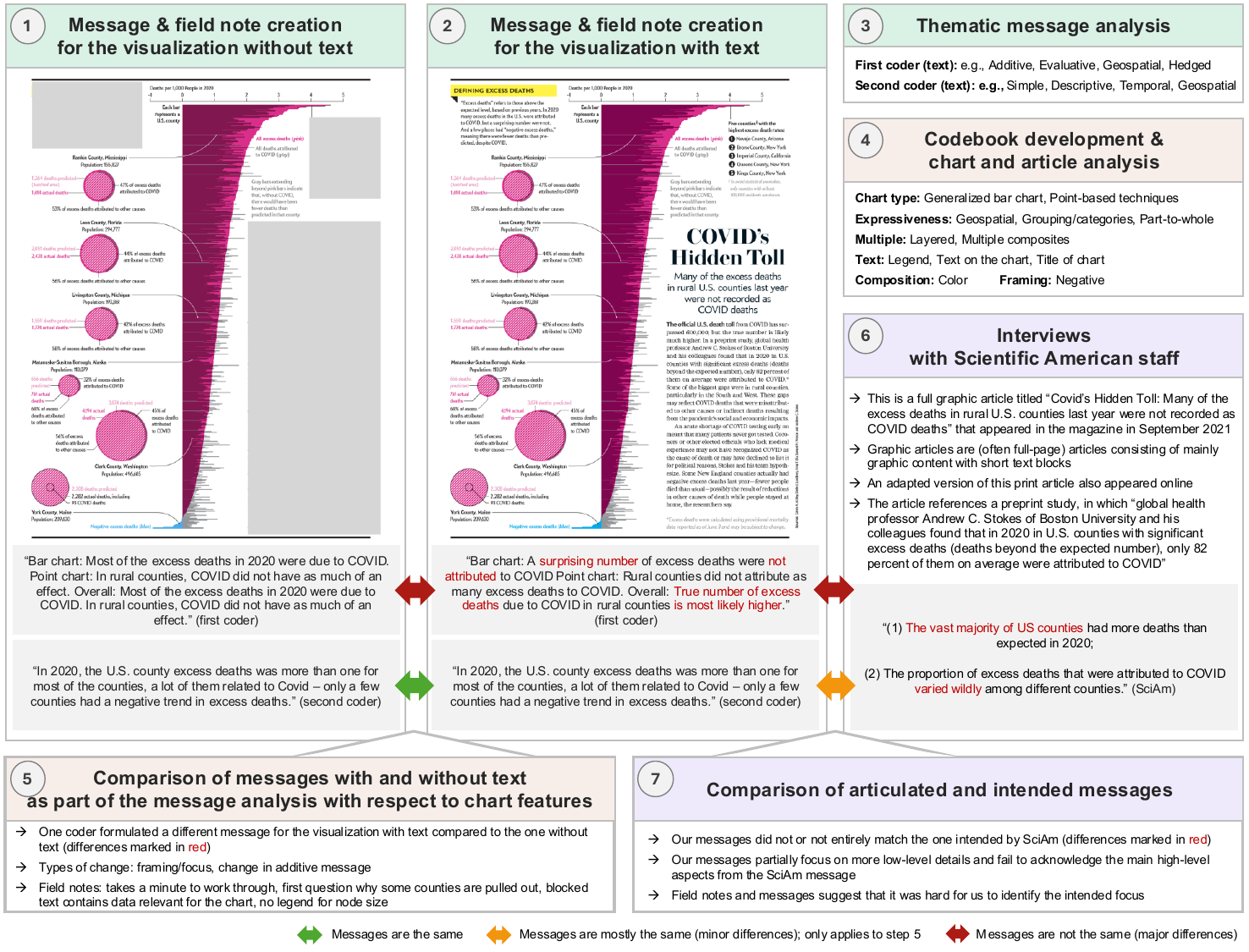}
    \caption{Case Study I illustrating the methodological process for an example chart (Vis CS-I), with a focus on message articulation and the comparison of messages across \textit{(no-text)} vs. \textit{(text)} conditions and with the intended message from Scientific American staff. The article reference can be found in the Scientific American Article Bibliography (S3, [26]) in the supplementary material.}
    \label{fig:casestudy}
\end{figure*}

To make the analytical process concrete, we walk through Case Study I in detail before presenting the full findings. Each visualization was interpreted independently by two coders. Case Study I illustrates a case in which only one of the two coders changed their message after seeing the chart with text. As \autoref{fig:casestudy} shows, the first coder's message shifted substantially: where they initially focused on the overall scale of COVID-related excess deaths, the addition of text led them to reframe their interpretation around the proportion of excess deaths not attributed to COVID and the likelihood of undercounting in rural counties. Their field notes indicate uncertainty about specific chart elements, specifically why certain counties were highlighted and how node size and texture should be interpreted. The second coder, by contrast, articulated the same message for both conditions, suggesting that the added text did not shift their interpretation. A similar tension between high-level and low-level interpretation emerged in the comparison between articulated and intended messages: the first coder's articulated messages focused on lower-level details, while the SciAm intended message emphasized higher-level patterns across counties. Design elements not intended as focal points may be interpreted as such when their purpose is unclear to the reader, underscoring the importance of transparency in design choices.

\subsection{Reader perspective: Which messages are articulated?} 
\label{subsec:findings-reader}
A systematic comparison of messages revealed recurring patterns in their formulation and expression. Through iterative thematic analysis, we developed a typology of 16 message types organized into six categories (\autoref{tab:message_typology}) that reflect variations in message granularity, structure, articulation, relations, inference, and content. Messages may be classified into multiple categories and types that are not mutually exclusive. The typology captures recurring ways in which messages are articulated, prioritizing thematic patterns over quantitative distributions. Across categories, messages consistently went beyond simple description, frequently incorporating interpretative, evaluative, and relational elements.

\textbf{Granularity.} Messages varied in their level of granularity, from general to specific. \textit{General messages} provided a high-level account of the visualization, with little detail. \textit{Mid-level messages} included some detail but were not very specific. \textit{Specific messages} conveyed precise information such as specific years, locations, values, or percentages, and sometimes described multiple parts of a chart rather than an overall trend. Developing more specific messages was often part of the sensemaking process, as coders refined their initial interpretations through closer engagement with chart details.

\textbf{Structure.} Messages varied in structure. \textit{Simple} messages expressed only a single insight, in contrast to \textit{additive} messages, which combined multiple sub-statements, often describing different parts of a chart or a series of small multiple charts. Additive structures frequently co-occurred with higher granularity, likely because complex or multi-part charts prompted coders to describe each component separately before synthesizing an overall message.

\textbf{Articulation.} We further identified variations in message articulation. \textit{Example-based} messages highlighted specific data points from the visualization and occurred most often in charts showing grouped or categorical data. In some cases, coders used examples when the amount of information in the chart exceeded what could be captured in a brief message, selecting the most prominent or top-ranked trend. \textit{Hedged} messages expressed uncertainty about the interpretation, either about what was being depicted or about the validity of the coder's reading of the data. These occurred more frequently in vis \textit{(no-text)}, where limited context made confident interpretation difficult. \textit{Chart-referential} messages appeared in a small minority of cases, where coders used chart-referential language to ground messages referring to small multiples or to describe an unfamiliar chart format. \textit{Source-referential} messages explicitly referenced the data source, study, or methodology underlying the visualization, such as naming specific models, datasets, or research findings.

\textbf{Relations.} In some cases, messages articulated relationships between data elements rather than focusing on single observations. \textit{Comparative} messages highlighted similarities or differences between categories, locations, or time periods. \textit{Causal or associative} messages expressed relationships between variables, either implying causation or describing patterns of co-occurrence.

\textbf{Inference.} Messages differed in how far they went beyond what is directly visible in the chart---a dimension that connects directly to our argument that communicative meaning cannot be reduced to decoding accuracy alone. \textit{Descriptive} messages restated observable patterns directly reflected in the visualization. \textit{Interpretative} messages extended beyond the displayed data by offering explanations or interpretations. \textit{Evaluative} messages went furthest, incorporating judgments, broader contextual framing, or personal perspectives that extended beyond the chart itself. \textit{Nonsensical} messages occurred exclusively in vis \textit{(no-text)}, arising when the visual encoding alone was insufficient to support any coherent interpretation.

\textbf{Content.} Content-wise, messages most commonly reflected patterns related to spatial location, temporal context, and change. Approximately half of all messages included \textit{geospatial} information (49\%), ranging from global patterns to local impacts; the latter appeared particularly in more recent charts addressing the socioeconomic effects of climate policies or the pandemic, such as effects on alcohol consumption or mental health. Notably, only about a quarter of charts encoded geospatial information visually (\autoref{fig:chart-features}), suggesting that much of the spatial context in messages was derived from titles, labels, or captions rather than from visual encodings. 54\% of messages referred to \textit{temporal} aspects, spanning both historical trends and future projections, such as specific years or broader time periods (\autoref{fig:chart-features}). The vast majority of messages included a description of \textit{change}, either explicitly or implicitly; those that did not emphasize change instead tended to compare categories. 

Given our focus on climate change and pandemics, the prevalence of geospatial, temporal, and change-related content reflects the nature of the corpus. However, the importance of these dimensions in data summarization has also been demonstrated beyond topic-specific contexts \citep{koesten2020everything}, suggesting broader relevance. Mentions of specific social groups or sectors occurred less frequently and were therefore not treated as a separate content category.

\begin{table*}[htpb]
\centering
\caption{Message typology with example messages indicating whether the message was articulated for vis \textit{(no-text)} or \textit{vis (text)}, and which coder (C1-C4) formulated the message}
\label{tab:message_typology}
\small
\begin{tabular}{>{\raggedright\arraybackslash}p{1.65cm} >{\raggedright\arraybackslash}p{2.6cm} p{10.9cm}}
\toprule
\textbf{Category} & \textbf{Type} & \textbf{Example message} \\
\midrule

\multirow{3}{*}{\textbf{Granularity}}
& General 
& {``Precipitation is changing globally -- in some places, it is getting drier, in others wetter.'' (vis \textit{(no-text)}, C1)} \\
\cmidrule(lr){2-3}

& Mid-level 
& {``Extreme weather events are becoming more frequent, particularly heat waves. Exposure will be much higher for younger generations and will vary according to world region and country income.'' (vis \textit{(text)}, C1, Vis A in \autoref{fig:examplecharts})} \\
\cmidrule(lr){2-3}

& Specific 
& {``Prediction of the risk of Malaria transmission will multiply in several parts of the earth by 2020 (relative to the average risk in 1961-1990), taking into account a temperature increase of about 2 degrees Fahrenheit. Especially in Europe, Asia, and small parts of the US, the risk of transmission might double, whereas the risk in other areas (e.g., Africa) might also increase but by a lower factor.'' (vis \textit{(text)}, C4)} \\

\midrule

\multirow{2}{*}{\textbf{Structure}}
& Simple
& {``The frequency of flood events at a given height in Atlantic City is projected to increase as baseline sea level grows.'' (vis \textit{(text)}, C3)} \\
\cmidrule(lr){2-3}

& Additive 
& {``The temperature [...] decreased every year until approx. 1980; the rising trend indicates that the temperature is steadily increasing from this point on. The global sea level [...] decreased every year until approx. 1980; the rising trend indicates that the global sea level is steadily increasing from this point on. The northern hemisphere snow cover [...] increased nearly every year until approx. 1980; the falling trend indicates that the snow cover is decreasing from this point on. Message for entire graphic: While temperature and global sea level is rising, northern hemisphere snow cover is decreasing.'' (vis \textit{(text)}, C4, Vis G in \autoref{fig:examplecharts})} \\

\midrule

\multirow{3}{*}{\textbf{Articulation}}
& Example-based
& {``Causes of deaths per week by different illnesses in the US, in relation to COVID -- where heart disease and cancer show a consistently high rate of over 10,000 per week (heart disease being higher).'' (vis \textit{(no-text)}, C1)} \\
\cmidrule(lr){2-3}

& Hedged  
& {``Areas with no change are in Central America and India. (Possibly where malaria is already high)'' (vis \textit{(no-text)}, C2)} \\
\cmidrule(lr){2-3}

& Chart-referential 
& {``In the scenario represented by both the red and black line, there are lower levels of $CO_2$ at more southern [...] and higher levels at northern latitudes.'' (vis \textit{(no-text)}, C2)} \\
\cmidrule(lr){2-3}

& Source-referential  
& {``With rising emissions, the temperature on earth will also rise [...]. This is based on 22 models from 17 different programs.'' (vis \textit{(text)}, C4)} \\

\midrule

\multirow{2}{*}{\textbf{Relations}}
& Comparative
& {``A plant-based diet produces less greenhouse gases than eating animal-based foods.'' (vis \textit{(no-text)}, C3)} \\
\cmidrule(lr){2-3}

& Causal / Associative 
& {``Lockdowns led to an unprecedented and sustained drop in flu cases.'' (vis \textit{(text)}, C2)} \\

\midrule

\multirow{4}{*}{\textbf{Inference}}
& Descriptive
& {``Winter sea ice has decreased since 1979 [...]. Winter air temperature, water vapor, and arctic amplification have increased at the same time.'' (vis \textit{(text)}, C1)} \\
\cmidrule(lr){2-3}

& Interpretative 
& {``The greenhouse effect has delayed a natural glaciation period, which is likely to happen once we stop making the planet warmer.'' (vis \textit{(text)}, C1)} \\
\cmidrule(lr){2-3}

& Evaluative 
& {``A surprising number of excess deaths were not attributed to COVID.'' (vis \textit{(text)}, C2)} \\
\cmidrule(lr){2-3}

& Nonsensical
& {``Showing something on planet Earth.'' (vis \textit{(no-text)}, C1)} \\

\midrule

\multirow{3}{*}{\textbf{Content}}
& \parbox[t]{\linewidth}{Geospatial}
& {``China and the US have the highest carbon dioxide emissions, with Guangzhou and New York being the cities with the highest domestic emissions.'' (vis \textit{(text)}, C3)} \\
\cmidrule(lr){2-3}

& \parbox[t]{\linewidth}{Temporal}
& {``Carbon dioxide has been increasing between 1958 to 1969.'' (vis \textit{(text)}, C1)} \\
\cmidrule(lr){2-3}

& Change-oriented 
& {``Daily helpline calls have increased for the 6 weeks of the pandemic outbreak, when they started to decrease again.'' (vis \textit{(no-text)}, C1, Vis C in \autoref{fig:examplecharts})} \\

\bottomrule
\end{tabular}
\end{table*}

\paragraph{Summary} Taken together, the typology reveals that articulated messages extend well beyond simple description. The prevalence of interpretative, evaluative, and relational types, alongside the emergence of nonsensical messages exclusively in the no-text condition, underscores both the complexity of visualization sensemaking and the critical role of textual elements in enabling coherent message articulation. These patterns are examined further from the chart perspective in the following section.

\subsection{Chart perspective: What role do chart features play?}
\label{subsec:findings-chart}
From a chart perspective, we analyzed the visualizations to identify patterns in visual data communication and to understand how chart features relate to sensemaking and articulated messages. We first describe recurring chart types and design characteristics across the corpus (Section~\ref{subsec:findings-chart-characterization}; \autoref{fig:chart-features}). We then examine how sensemaking processes were influenced by these features (Section~\ref{subsec:findings-chart-messagechart}) and how takeaway messages changed with the addition of textual elements on the visualizations (Section~\ref{subsec:findings-chart-messagetext}).

\subsubsection{Which chart types and design characteristics are used?}
\label{subsec:findings-chart-characterization}
We first examine chart features in the corpus, namely chart types and design characteristics such as color encoding and grouping. \autoref{fig:chart-features} provides a summary of the frequency of these chart features in the corpus.

\textbf{Chart types.} We characterized chart types according to \citet{chen2022not}, allowing charts to be assigned to multiple categories. Nearly two-thirds of the charts in the corpus were line-based visualizations (64\%), ranging from simple trend lines to more complex uses of lines, such as in maps or to indicate reference points in time. Other chart types included generalized area charts such as maps (31\%), point-based charts (20\%), generalized bar charts (14\%), and continuous color-encoded charts (12\%). Chart types suggest that most visualizations depict quantitative data, whereas few (n=10) present categorical data. Over one-third of visualizations (37\%) combined multiple chart types. In addition, 43\% of charts were presented together within an article, either as small multiples, a series of similar charts using the same axes or scales, or as composite figures combining different chart types or scales (see Vis A in \autoref{fig:examplecharts} and Vis CS-I in \autoref{fig:casestudy}). This compositional complexity likely contributed to the prevalence of additive message structures observed in the reader perspective.

\textbf{Chart characteristics.} We identified several recurring design characteristics across the corpus. Color was used in most visualizations to encode data (60\%), and 61\% used groupings to present data, usually adding context (e.g., ``counties with the most excess deaths due to COVID-19''). Geospatial data appeared in 24\% of charts, most commonly in area charts such as maps or line-based visualizations. Most charts included temporal data, either referencing historical time (55\%) or future predictions (24\%); note that these categories overlap, as some visualizations included both. In contrast, charts rarely included humanizing elements such as photographs or illustrations of people (5\%).

We also coded the visualizations for missing data, uncertainty, and text use. A small fraction of charts (6\%) indicated missing data directly on the chart via a label or visual representation, as in Vis D in \autoref{fig:examplecharts}. Representations of uncertainty were present in 15\% of charts, most commonly by indicating a range of possible values (see Vis E and G in \autoref{fig:examplecharts}). 54\% of the visualizations included text instances in the chart area beyond titles, with some including several types of textual annotations. This text more often provided explanations, ranging from brief sentences to longer passages describing chart content, highlighting data, or explaining how to read the visualization, than references to data sources. Finally, coders assessed each visualization's overall framing; as these were subjective codes that did not require consensus, they reflect individual coder perceptions rather than objective classifications. Most visualizations were perceived as neutral (60\%), while 36\% were seen as negatively framed and only 4\% as positively framed, possibly reflecting the severity of the topics covered.

\subsubsection{How do chart features influence sensemaking?}
\label{subsec:findings-chart-messagechart}
Chart types and design characteristics shaped our sensemaking and the interpretive effort they required. Factors influencing interpretation included visual encodings, visual complexity, numerical precision, visual emphasis, framing, personalization, and the alignment between charts and accompanying text. Field notes documenting these interpretation processes provide the primary evidence for the findings in this section.

\textbf{Visual encodings.} Color was frequently used to distinguish between data types, scenarios, and time points (see Section~\ref{subsec:findings-chart-characterization}), but some color choices created perceptual difficulties. In some cases, similar colors such as red and orange on a single line made differences hard to see and required closer inspection. While most of the corpus consisted of common chart types, some visualizations combined chart forms in ways that introduced additional complexity for sensemaking. Encountering unfamiliar chart combinations was at times challenging, as one coder noted:

\begin{quote}
    I struggled with this vis at first, as it looks a bit different than other visualizations I have seen, even though the elements and chart types are common (point and bar). After a minute, though, when I got the feel for how the pieces work together, I found that I really like the way the points or nodes provide an in-depth look at some of the counties. (Field note excerpt, C2, Vis CS-I) 
\end{quote}

\textbf{Visual complexity.} Visualizations with denser layouts or more cluttered designs made it harder to distinguish individual elements and trace relationships across the data. Charts that combined multiple time scales or presented both historical and future data required particular effort to reconcile:
\begin{quote}
     It's really difficult to balance and understand the different timescales in these charts. Vis 3 [...] has data from storms in 2002, while the data on the chart is for current and future levels. (Field note excerpt, C2)
\end{quote}

\textbf{Numerical precision.} Coders' sensemaking did not always rely on exact numerical values, as one-third of charts displayed no precise values (\autoref{fig:chart-features}), requiring visual estimation of quantities and encouraging coders to get a \textit{feel} for magnitudes or rely more heavily on textual annotations. Coders responded in different ways, sometimes approximating values and at other times expressing frustration:

\begin{quote}
    When writing the messages, I was also looking for trends and `visually averaging' the steepness of the lines for the different categories. I have no idea if this is what is an intended way of interpreting these charts; it feels a bit awkward and like a hack. (Field note excerpt, C2)

    Maybe a clearer hint about the 6 times more number would have been helpful in Vis 1, so one does not have to calculate it on their own. (Field note excerpt, C3) 
\end{quote}

\textbf{Visual emphasis.} Visual emphasis guided sensemaking by directing attention to particular values, categories, or relationships. Notably, emphasis sometimes emerged through omission rather than visual highlighting; e.g., a visualization showing countries with a carbon tax did not represent countries without such measures, and coders typically focused on the countries depicted. The accompanying text could also introduce emphasis that coders found misleading:

\begin{quote}
    In the summarizing information, it mentions the lowest and the highest (carbon) tax rate per metric ton of $CO_2$. I have to say the lowest is slightly misleading, in my opinion, because many countries do not pay any carbon tax, so it emphasizes the ones that are paying low rates in a negative way, instead of those that pay none. (Field note excerpt, C1)
\end{quote}

\textbf{Framing.} Sensemaking was also influenced by the perceived framing or tone of a visualization, which often emerged through color choices, captions, or title wording. Earlier article and chart titles tended toward more neutral or objective formulations, e.g., `Tree Rings and Climate' (1972) or `Carbon Dioxide and World Climate' (1982), while more recent titles carried greater emotional appeal, such as `Too Sunny in Philadelphia? Satellites Zero in on Dangerous Heat Islands' (2017) or `The Very Real Death Toll of COVID-19' (2021). Field notes also suggest a shift in caption practices: older captions more often included technical detail such as data collection methods, uncertainty representations, or legend-like guidance, whereas more recent captions more frequently conveyed broader narrative context. The framing introduced by text could substantially alter interpretation, as one coder observed:

\begin{quote}
    Looking at the charts alone, I thought this was a very negative framing---look where we are headed as we are moving inextricably toward a 2-degree increase in temperature. With the caption, the chart takes on a positive outlook---look what we can avoid by being more ambitious in our climate policy! (Field note excerpt, C2)
\end{quote}

\textbf{Personalization.} While only a small number of visualizations incorporated humanizing elements, locally relevant references or accompanying photographs introducing a human perspective appeared in several cases. These elements could foster personal relevance and deeper engagement, but their connection to the data was not always clear, leading to mixed responses:

\begin{quote}
    This is the first time that I have seen scenarios labeled this way (in terms of the president who is ultimately responsible for a policy) [..] I googled the exact years of Obama’s presidency [..] I did additional research and made connections between policy and $CO_2$ data, maybe because this is a bit close to home for me. (Field note excerpt, C2)

    I’m also not sure if I get the story of this person in the yellow coat (what is the connection to the chart parts?). (Field note excerpt, C3)
\end{quote}

\textbf{Chart-article alignment.} Sensemaking often involved relating multiple visualizations within an article to one another. Small multiples and composite charts, which were common in the corpus, helped coders work through complex concepts, trace patterns over time, and identify anomalies or missing data. At times, however, large numbers of charts within a single article increased interpretive load. Interpretation also became more difficult when the focus or level of detail differed between charts and accompanying text:

\begin{quote}
    The main informative factors are the small headings next to the individual charts, which tell a coherent story. However, they do not fully fit the summarizing info on top of the charts. (Field note excerpt, C1)
\end{quote}

\subsubsection{How do textual elements change articulated messages?}
\label{subsec:findings-chart-messagetext}
For each visualization, we compared messages articulated for vis \textit{(no-text)} and vis \textit{(text)} to assess whether and how textual elements beyond structural labels, such as explanatory annotations or titles, shaped the articulated message. Field notes document the difficulties coders encountered when interpreting charts without text. When writing initial messages, coders relied on any retained textual cues to piece together a coherent interpretation. Even when all chart text was available, additional explanation was sometimes sought in the accompanying articles. This was particularly the case for older visualizations, e.g., those published in the 1970s, where design elements were more limited or under-explained.

\begin{quote}
   For most [vis \textit{(no-text)}] I have no idea what the chart is trying to say. I am mostly connecting the few textual annotations on the chart and trying to make a comprehensive sentence out of them. (Field note excerpt, C1)
\end{quote}

\textbf{How frequently do messages change?} Over two-thirds of messages articulated for vis \textit{(text)} differed from their corresponding messages for vis \textit{(no-text)} (68\%). In 27\% of cases, messages remained unchanged even with the addition of textual elements, and in 6\%, only minor changes occurred, such as adding a word or short phrase. The proportion of messages that changed varied across chart types. For line-based visualizations, general area charts, and continuous color-encoded charts, over 60\% of messages differed between conditions. Messages for bar charts (45\%) and point-based charts (49\%) changed less frequently. By contrast, messages for infographics and schematic representations differed in over 80\% of cases; however, these categories were less frequent in the corpus and should therefore be interpreted cautiously.

\begin{figure}[htbp]
    \centering
        \includegraphics[width=\textwidth]{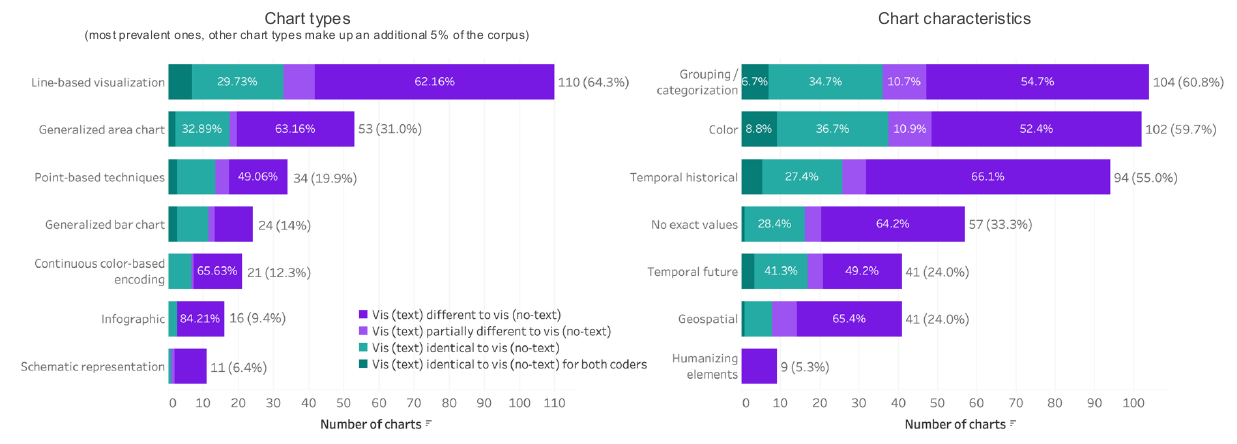}
    \caption{Distribution of chart types (left) and chart characteristics (right) in the corpus and differences between messages articulated for vis \textit{(no-text)} and vis \textit{(text)}. Bars show, for each category, the proportion of messages that were different, partly different, identical, or identical across both coders when text was added. Counts and percentages indicate the overall frequency of each category in the corpus. Percentages within bars are calculated per category. Only the most frequent categories are shown; additional chart types include generalized matrix/grid (7, 4.1\%), text-based encoding (1, 0.6\%), and graphs/networks/meshes (1, 0.6\%). Note that a single visualization may be coded as multiple chart types.}
    \label{fig:chart-features}
\end{figure}

Chart characteristics showed no consistent influence on message change rates; however, some patterns are worth noting (S7). Charts displaying future temporal information showed the lowest proportion of differences (49\%); however, this characteristic occurred less frequently in the corpus. Historical temporal information, geospatial charts, and charts without exact values showed somewhat higher change rates (approximately 64--66\%), while grouping and color use were close to the overall average (approximately 52--55\%). Because chart characteristics often co-occur, these patterns should be interpreted cautiously. \autoref{fig:chart-features} shows an overview of the distribution of chart types and characteristics in the corpus, as well as differences between messages articulated for vis \textit{(no-text)} and vis \textit{(text)}.

In a small number of cases, messages of vis \textit{(no-text)} and vis \textit{(text)} contradicted each other, either because the visual encoding alone led to a different interpretation, or because the broader message only became apparent with textual context. Examples comparing messages of vis \textit{(no-text)} and vis \textit{(text)} for the charts in \autoref{fig:examplecharts} are provided in S6.T1; additional details on change statistics are shown in \autoref{fig:chart-features}.

\textbf{What types of differences occur?} Beyond how frequently messages changed, we also examined the nature of those differences. To better understand how textual elements shape interpretation, we identified three broad categories of change between messages of vis \textit{(no-text)} and vis \textit{(text)}: changes related to i) granularity, ii) types of detail, and iii) interpretative emphasis. Example messages illustrating these change types are shown in \autoref{tab:message_changes}.

\textit{Granularity.} Messages of vis \textit{(text)} were typically more detailed than their vis \textit{(no-text)} counterparts, a pattern reflected in message length: vis \textit{(no-text)} messages had an overall median length of 145 characters, whereas vis \textit{(text)} messages were longer (median 187 characters), with variation across coders (S6.T2). In some cases, however, vis \textit{(text)} messages became more concise, as illustrated in the Granularity example in \autoref{tab:message_changes}, where the addition of text prompted a shift toward a higher-level, more general message rather than a more detailed one.

\textit{Types of detail.} Messages also changed in the types of detail included. Many messages were revised to incorporate scientific terminology or jargon, often taken directly from captions or other textual elements. In some cases, messages of vis \textit{(text)} included additional details about how the data were created, most visibly in visualizations comparing different data sources or models. Geospatial or temporal information was often added, though in some cases it was also reduced, possibly reflecting a move toward more abstract interpretation. Other messages of vis \textit{(text)} included additional concrete reference points to place the message in the context of expected or abnormal values.

\textit{Interpretative emphasis.} Interpretative emphasis also shifted between conditions. The focus of some messages shifted to highlight different parts of a visualization or dataset, particularly for charts with many groupings or categorical data. In other message pairs, framing shifted, with certain elements qualified through evaluative language such as `surprising', `concerning', or `fragile' (see evaluative message type, Section~\ref{subsec:findings-reader}).

\begin{table*}[htpb]
\centering
\caption{Types of changes between message of vis \textit{(no-text)} and message of vis \textit{(text)} with example messages}
\label{tab:message_changes}
\small
\begin{tabular}{>{\raggedright\arraybackslash}p{1.8cm} p{6.6cm} p{6.3cm}}
\toprule
\textbf{Change} & \textbf{Message of vis \textit{(no-text)}} & \textbf{Message of vis \textit{(text)}} \\
\midrule

\textbf{Granularity}
& ``Reported lives lost from Covid per week peaked around the beginning of 2021, with 100{,}000 deaths per week, mostly in Europe and the Americas.'' (C3)
& ``The pandemic has caused an enormous loss of life across all parts of the world.'' (C3, Vis B in \autoref{fig:examplecharts}) \\
\midrule

\textbf{Type of detail}
& ``The most calories per square meter are used/produced at around 10 degrees South (latitude), the least at around 20 degrees North.'' (C1)
& ``The amount of heat lost through evaporation and gained through precipitation at each latitude.'' (C1) \\
\midrule

\textbf{Interpretative emphasis}
& ``The ice sheet in most of Antarctica has been relatively stable for the last 15 million years; only a small portion of Antarctica [...] is certain to have been impacted by global warming.'' (C2)
& ``West Antarctica is the most fragile and has disintegrated rapidly; we do not know for sure if this is due to climate change.'' (C2) \\

\bottomrule
\end{tabular}
\end{table*}

\textbf{How does text alter the sensemaking process?} Textual elements not only changed the articulated messages; they also altered how coders arrived at those messages. When text was present, it often served as a source of disambiguation and guidance, prompting revision of how the data were interpreted and meaning was constructed. These observations point to shifts in the sensemaking processes underlying the message differences described above. We observed three recurring process-level adjustments: abstraction, explanatory reasoning, and recontextualization.

Through \textit{abstraction}, textual cues encouraged coders to move from detailed observations to higher-level interpretations of underlying patterns or phenomena, connecting to the granularity shifts described above. Instead of focusing on individual values or regions, coders synthesized broader trends and overarching implications. Through \textit{explanatory reasoning}, text prompted interpretation of relationships, mechanisms, or implications not immediately evident from the visual encoding alone. This included clarifying units of measurement, identifying thresholds, or reasoning about causal mechanisms and future consequences. In some cases, textual elements supported understanding of more complex interactions between variables, such as the interplay of different aspects of global warming. Through \textit{recontextualization}, text led coders to reconsider framing and relevance by connecting the visualization to broader contexts. Captions, annotations, or accompanying text sometimes shifted interpretative emphasis, directed attention to particular aspects, or prompted connections to prior knowledge or information in the surrounding article.

\paragraph{Summary} Textual elements played a decisive role in shaping articulated messages: over two-thirds of messages differed between conditions, with changes spanning granularity, types of detail, and interpretative emphasis. These differences reflect deeper shifts in the sensemaking process, from abstraction and explanatory reasoning to recontextualization, highlighting text as a key mediator between visual encoding and communicative meaning. The following section examines how these articulated messages relate to the messages intended by visualization producers.

\subsection{Producer perspective: How are messages intended?}
\label{subsec:findings-producer}
This section presents the final perspective in our message analysis: that of the chart producers at Scientific American. We examine how messages are formulated in production workflows, how they are shaped by audience and communication goals, and how visualizations are situated between narrative context and standalone readability (Section~\ref{subsec:findings-producer-planning}). We then compare producers' intended messages with a sample of the vis \textit{(text)} messages we articulated (Section~\ref{subsec:findings-producer-comparison}). Participants are referred to as P1--P11 in this section. A detailed analysis of the interviews from a complementary perspective is available by \cite{gregory2024data}.

\subsubsection{How are messages planned and situated in visualization practice?}
\label{subsec:findings-producer-planning}

\textbf{Formulating messages in production workflows.} Scientific American staff described message development as an early step in visualization production. Messages, sometimes described as goals (a term that itself reflects the close relationship between communicative intent and design purpose), were often articulated at the outset of a project, either informally or in writing, to clarify the purpose of a graphic and guide subsequent design decisions. These planned messages could be clear and concise, where \textit{``the goal of this graphic is to X, Y, and Z''} (P1), or involve detailed elements intended to support broader takeaways. In some cases, however, the message was not fully defined at the start. When working with unfamiliar or complex datasets, producers described using exploratory visualization to identify patterns and allow a story to emerge:

\begin{quote}
    More often than not, I do have a clearer sense of what the goal is before I start working. I guess the only time when that might not be the case is if it's a really large data set that maybe I haven't seen visualised. And so I have to sort of like, do a little bit more exploration and playing with the visualisation, and then kind of a story emerges. (P2)
\end{quote}

Message planning was also described as a collaborative and editorially negotiated process. Early articulation helped streamline internal conversations among editors, writers, artists, scientists, and designers. Visualization producers frequently worked alongside writers and editors who \textit{``have a clear sense of what they want the graphic to show''} (P2), and aligning these perspectives helped ensure that the visualization supported the article's narrative. Text also frequently informed the development of graphics, e.g., when designers began with a complete written story and needed to consider which \textit{``points we want to hit in words only''} (P4). Reaching an agreement on the intended message before design work began streamlined workflows and reduced unnecessary revisions:
\begin{quote}
    At the beginning, it was this is what we want to say, these are the points we want to hit -- in words only. And then [...] if we agree that's the message that we want to give, then I would go and sketch out some ideas and share it with the editors. (P4)
\end{quote}

\textbf{Shaping messages through audience and communication goals.} Message planning was described as shaped by a strong sense of responsibility toward readers and broader public communication goals. Participants emphasized that visualizations are created to support readers' understanding and that intended messages are considered from the reader's perspective:
\begin{quote}
    ``We're not doing this for ourselves. We're doing it for the reader. So we were always trying to think of how the reader will perceive this. We sometimes succeed and sometimes don't.'' (P3)
\end{quote}

Some participants described how attention to message planning intensified during the COVID-19 pandemic, as they felt they could no longer assume shared background knowledge or trust in scientific information. Participants described becoming more attentive to how messages might be interpreted or misinterpreted by diverse audiences:
\begin{quote}
    I think a lot more about the message and [...] how it could be misinterpreted. Like before the pandemic, [...] I was pretty sure [the readership] already agreed with, or was open to, the scientific method and was kind of on board with what I was saying. Whereas now, especially with covering COVID, I can't assume that the audience is necessarily going to agree and understand everything that I'm saying. (P1)
\end{quote}

This reader-oriented approach to message planning also shaped how visualizations were designed to meet different needs and encourage engagement. Participants described ongoing efforts to present complex topics such as climate change in ways that resonate with diverse audiences and increase their impact:
\begin{quote}
  We have been showing that kind of information for two decades, at least in different forms [..] That is an area where we are trying to meet different people's needs in different ways, in part because there is such a sense of urgency [...] for 20 years, these charts have not been making enough of an impact. How else can we present this information in a way that connects with people? (P1).  
\end{quote}

\textbf{Situating visualizations in narrative context.} Scientific American staff described using visualizations and text together to serve a set of purposes, e.g., explaining, stepping readers through an article, or pointing out uncertainties to nuance the seeming objectivity of data. Participants recognized that visualizations never truly exist independently, as they are \textit{`talking about and referencing a very particular thing'} and only make sense within a broader story (P6). Visualizations can appear self-contained only when essential context is embedded within the graphic itself, for example, through text or narrative placement (P2, P6). Placement and contextualization within the article, therefore, matter considerably, as one participant noted:
\begin{quote}
    It's important to put those numbers in context. [...] But it's a challenge because I know we do lose readers, and as a news consumer myself, you know, I don't always make it to the bottom of the news story. And sometimes you have a really important message you want to get across, lower down, but you might lose people before they even get to that. (P8)
\end{quote}

At the same time, some participants described striving to make graphics as self-contained as possible, particularly anticipating circulation beyond the article context (P2, P4, P6). Producers considered how graphics might be interpreted in isolation and made design decisions to prevent messages from being taken out of context or misused:
\begin{quote}
    My plan was, even if you didn't read the article, you could still get the info from just working through the graphics and the photos and the captions (P4).
\end{quote}

This tension between context-dependence and self-containedness is a recurring design challenge, with producers navigating competing demands based on the intended audience, publication format, and the anticipated reach of the graphic.

\subsubsection{How do articulated messages relate to producers' intentions?}
\label{subsec:findings-producer-comparison}
We acquired intended messages for 18 charts from a senior Scientific American staff member, who shared internal documentation prepared during the production process. For this sample, we compared the producer's intended message with the messages of vis \textit{(text)} articulated by the research team. While we do not assume that charts convey a single fixed message, comparing communicative intent and interpretation allows us to examine where they align and where they diverge.

We observed varying degrees of alignment between intended and articulated messages, assessed in terms of central communicative emphasis rather than wording or level of detail. In 45\% of cases, the articulated messages conveyed the same central takeaway as the intended messages. In 38\% of cases, messages were partially aligned, capturing elements of the intended meaning while omitting nuance or reframing aspects of the message. In the remaining 17\%, the articulated messages diverged substantially, reflecting a shift in core communicative emphasis. \autoref{tab:intended_vs_articulated} shows exemplary message pairs for comparison.

\begin{table*}[htb]
\centering
\caption{Comparison of exemplary articulated messages with intended messages provided by Scientific American staff.}
\label{tab:intended_vs_articulated}
\small
\begin{tabular}{>{\raggedright\arraybackslash}p{2.6cm} p{6.8cm} p{5.8cm}}
\toprule
 & \textbf{Exemplary articulated message of vis \textit{(text)}} & \textbf{Exemplary intended message} \\
\midrule

\textbf{Major differences} 
&
``Bar chart: A surprising number of excess deaths were not attributed to COVID. Point chart: Rural counties did not attribute as many excess deaths to COVID. Overall: True number of excess deaths due to COVID in rural counties is most likely higher.'' (C2)
&
(1) The vast majority of US counties had more deaths than expected in 2020;
(2) The proportion of excess deaths attributed to COVID varied widely among different counties. (SciAm)
\\

\midrule

\textbf{Minor differences} (Vis F in \autoref{fig:examplecharts})
&
``Cycles in cooling and warming in different parts of the world have been normal, with certain parts of the world being warmer and certain parts being cooler. The current warming trend (since 1950) is not normal and is warmer everywhere.'' (C2)
&
``Global warming is not part of natural climate variability. Yes, global temperatures have fluctuated over time. But since around 1950, almost all years have been overwhelmingly warmer than a 2000-year average.'' (SciAm)
\\

\midrule

\textbf{Same takeaway} (Vis B in \autoref{fig:examplecharts})
&
``The pandemic has caused an enormous loss of life across all parts of the world.'' (C3)
&
``Lots of lives were lost very quickly around the world due to a virus.'' (SciAm)
\\

\bottomrule
\end{tabular}
\end{table*}

Overall, messages articulated by the research team were more likely to include specific examples, particularly concrete numbers such as years or high values, or the top categories shown on a chart, as in Vis B in \autoref{fig:examplecharts}: \textit{``Covid related deaths reported to the WHO reach up to 100,000 lives lost per week in the beginning of 2020''} (C1), compared with \textit{``Lots of lives were lost very quickly around the world due to a virus''} (SciAm). We hypothesize this reflects our role as researchers trained to present data as evidence rather than as narrative. Messages articulated for both conditions were more focused on the data itself, whereas producers' intended messages were expressed in higher-level language, focusing on what the data means when generalized. This may also reflect a bottom-up sensemaking strategy, in which coders are anchored in specific data points before synthesizing a broader message.

Often, differences between intended and articulated messages arose from focusing on different aspects of a chart. While one coder interpreted Vis CS-I in \autoref{fig:casestudy} similarly to the intended message, the other coder focused on details that diverged from the producer's intended focus. Field notes suggest that such differences occurred when coders could not determine which part of a chart was most important and therefore struggled to abstract the overall message. Charts with many details were more likely to be interpreted in accordance with the intended message when interpretation aids were present, such as a summary text or visual cues indicating relevance. Interpretation was also more difficult for charts with multiple intended messages, with coders sometimes failing to capture one of the key points. Taken together, these comparisons suggest that differences between intended and articulated messages most often reflect variation in emphasis, abstraction, and the availability of visual cues indicating relevance, rather than fundamental misunderstanding.

\paragraph{Summary} Visualization messages at Scientific American were planned early, shaped collaboratively, and oriented toward diverse audiences, with producers navigating a recurring tension between context-dependence and self-contained\-ness. Comparing intended and articulated messages revealed varying degrees of alignment, with differences most often reflecting variation in emphasis and abstraction rather than fundamental misunderstanding.
\section{Discussion}
\label{sec:discussion}
This work has examined visual data communication in Scientific American from three perspectives---those of readers, charts, and producers---to better understand how data on climate change and, to a lesser extent, pandemics are presented and interpreted. From the reader's perspective, we systematically articulated takeaway messages and accompanying field notes for each visualization. This process allowed us to analyze what constitutes a visualization message and how messages differ in structure and content. From the chart perspective, we examined how visual features and textual elements shape sensemaking by comparing interpretations of visualizations with and without accompanying text. From the producer's perspective, we interviewed Scientific American staff to examine intended messages and message planning practices. Comparing a sample of our articulated messages with those intended by chart producers highlighted the diversity of messages that can be interpreted from the same chart, which we hypothesize increases when visualizations address broad audiences.

We now bring together and discuss our findings across five key aspects: takeaway messages as an analytic lens on sensemaking (Section~\ref{subsec:discussion-reader}), the role of chart features including textual elements (Section~\ref{subsec:discussion-chart}), a reflection on intent versus interpretation (Section~\ref{subsec:discussion-producer}), our methodological contribution (Section~\ref{subsec:discussion-methodcontribution}), and the limitations of this work (Section~\ref{subsec:discussion-limitations}).

\begin{figure}
    \centering
    \includegraphics[width=0.9\linewidth]{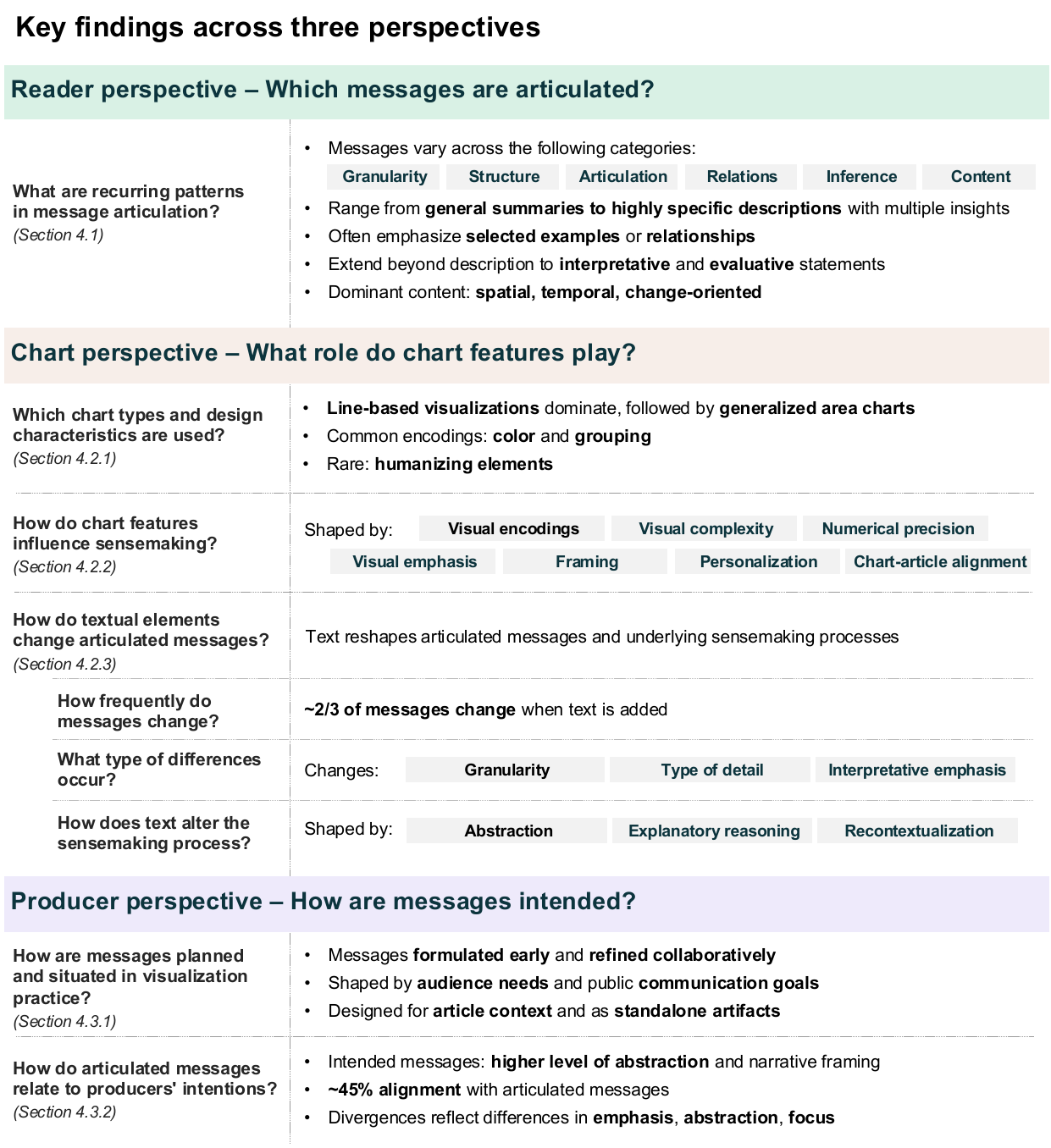}
    \caption{Summary of key findings across three perspectives, showing how messages are articulated, shaped by chart features and text, and relate to producers’ intentions.}
    \label{fig:placeholder}
\end{figure}

\subsection{Takeaway messages as an analytic lens on sensemaking}
\label{subsec:discussion-reader}

Making sense of data visualizations is an inherently complex process: it requires not only making sense of the underlying data but also interpreting how that data has been visually encoded and contextualized \citep{koesten2021talking, victorelli2020understanding, dourish2017stuff}. Our study adds evidence to a growing body of literature documenting the complexities of understanding visualizations \citep{lee2016how,pohl2017sense-making,baker2009using}. Our findings show that message articulation is an active sensemaking process rather than a mere formulation of facts. This process is shaped by factors related to the reader (e.g., familiarity with the topic and personal relevance), the chart (e.g., visual characteristics, on-chart text, article context), and the producer (e.g., intended message, design choices, framing). This situated meaning-making produced a wide range of articulated messages that varied not only in content but also in structure and interpretation. These observations challenge approaches that imply charts communicate meaning directly, for example, by focusing primarily on the effects of visual encodings on accuracy and task performance.

We drew on a thematic analysis of articulated messages to develop a typology comprising six message categories and 16 message types across visualizations on climate change and pandemics. While prior work has categorized messages in COVID-19 visualizations by topic and visualization type \citep{zhang2021mapping}, our typology takes a more generalizable approach by examining dimensions beyond content, such as granularity, structure, articulation, relations, and inference. Following \citet{koesten2021talking}, our aim was to analyze how messages emerge through individual sensemaking and to examine what is `taken away' from a chart and how. The resulting typology provides a vocabulary for describing how messages differ, not merely whether they are correct, supporting analyses of interpretation beyond accuracy metrics and enabling comparison across contexts.

Our typology resonates with the work of \citet{lundgard2022accessible}, who classify semantic levels in visualization descriptions to comprehensively capture chart content. Despite differing aims, there are notable similarities between the resulting typologies. Like their lowest semantic level, we identify messages that describe chart elements; however, such descriptions proved less informative for our goal of surfacing sensemaking. Instead, most articulated messages operated at the higher semantic levels described by \citet{lundgard2022accessible}. We also observed parallels with the typology introduced by Yang et al. \citep{yang2023explaining}. While their work focuses on verbal explanations of visualizations, aspects of their categorization overlap with our message typology, particularly in our `message articulation' category: the messages in our study rarely relied on concrete examples, a pattern also observed by \citet{yang2023explaining} among participants with high visual literacy.

The typology we propose provides a structured vocabulary for discussing the messages communicated through visualizations and can support more intentional communication design. Understanding the kinds of messages viewers derive from visualizations can inform how charts are planned and adapted for different audiences and help teams articulate communicative intent before design begins. Beyond human-centered workflows, the typology also has potential for computational applications. For example, it could inform automated caption generation by accommodating different levels of granularity or the inclusion of specific examples, and it may support approaches that extract semantic meaning from chart images \citep{wu2010recognizing,qian2021generating}. Message types could also guide content selection in template-based systems or inform feature weighting, as suggested by \citet{lundgard2022accessible}. Nevertheless, the typology is grounded in our corpus and may require adaptation when applied to other data domains or topics.

\subsection{How chart features and text shape interpretation in practice}
\label{subsec:discussion-chart}

Our chart corpus was dominated by line-based visualizations and temporal reasoning; approximately half of the articulated messages included geospatial information. This reflects both the topics in our corpus and the importance of spatial and temporal dimensions for making sense of data, as documented in prior work on dataset summarization \citep{koesten2020everything}, data search \citep{neumaier2019enabling}, and storytelling with data \citep{erete2016storytelling}. While we found little evidence within the charts themselves of techniques commonly used to evoke emotion or attention, such as humanizing elements, semantic icons, or expressive color use \citep{ferreira2024connecting,alebri2014visualizations,bartram2017affective,macedomorais2021can}, our findings suggest that affective framing is instead conveyed through surrounding elements, including photographs, accompanying text, or bespoke illustrations, which help underline a message.

While chart types and characteristics might influence what readers take away from visualizations, our findings suggest that the broader environment and framing in which charts are presented are critical for interpretation. In several cases, alignment between the chart and the article context was essential for understanding, underscoring that charts do not exist in isolation. This context-driven perspective also encompasses subjective reactions that shape engagement with visualizations \citep{kennedy2018feeling,cawthon2007effect}. Both our field notes and interviews reflect readers' personal responses, including emotional reactions, responses to visual presentation, and perceived rhetorical appeals such as topic framing. Our field notes further reveal a particular appreciation for detail and composition, highlighting how visual complexity and appeal influence cognitive effort and reception. Coordinating multiple design decisions to guide readers' attention in this way can be understood as a form of craft. As with physical objects, the care and skill evident in creation can motivate engagement, as carefully crafted artifacts are perceived as more valuable and deserving of attention \citep{norman2004emotional}.

In many cases, separating additional textual elements from charts was impossible while preserving meaning. The presence of text prompted shifts in sensemaking processes, including changes in abstraction, explanatory reasoning, and recontextualization. In line with \citet{kim2021towards}, we find that annotations, captions, and titles associated with charts and articles are crucial for engaging readers. These elements set readers up for `sensemaking success' by highlighting relevant data, explaining thresholds, and providing guidance on how to read a chart. This observation aligns with literature calling for improved alternative text descriptions of charts \citep{jung2022communicating} and with recent work examining the integration of charts and captions \citep{stokes2023role}.

Overall, these findings highlight that meaning emerges from the interplay of multiple chart features and contextual elements, reinforcing the view that design choices should be treated as consequential communicative and narrative decisions. Because annotations and titles play a critical role in guiding interpretation, future work and tools should support designers in crafting effective textual additions. At the same time, visual complexity should be considered in relation to the textual support available, as our findings suggest that complexity becomes particularly problematic when textual elements are insufficient to guide interpretation. Design tools could assist this process by generating adaptable text suggestions aligned with communicative goals or by providing visual complexity checks that flag potential cognitive load issues.

\subsection{Intent versus interpretation in popular science visualization} 
\label{subsec:discussion-producer}
Our findings suggest that staff at Scientific American carefully consider design choices, including chart types, data encodings, and the interplay between visual and textual elements across an entire article (cf. \cite{gregory2024data}). Some choices reflect perceptual research and established guidelines or conventions (see \citet{franconeri2021science}). Others address audience engagement, producers' affective intentions, and aesthetic presentation \citep{he2022beauvis,lee-robbins2022affective}. The message of a visualization and its communicative goal are typically constructed collaboratively through negotiation among editors, writers, and visualization designers, and are centered on audience considerations \citep{schuster2026practitioners}. In this process, visualization producers treat charts as narrative devices with intentional communicative goals and planned takeaways. This supports perspectives that frame chart creation as a rhetorical practice \citep{prantl2026untangling} and aligns with broader developments toward data storytelling \citep{shao2024data}. Similarly, the role of emotion in visualization design is increasingly recognized \citep{kostelnick2016re-emergence,kennedy2018feeling}; in our findings, affective considerations shaped both design decisions and our sensemaking processes (cf. \cite{wang2019emotional}).

Our findings underline that visualizations always stand in relation to surrounding content. In the growing field of online data communication, helping readers navigate visualizations to support sensemaking is paramount. Research on narrative visualization~\citep{segel2010narrative}, storytelling and comics \citep{bach2017emerging}, and interactive charts~\citep{firat2022interactive} moves in this direction. However, we argue that a sustained focus on context in relation to readers' sensemaking and interpretation remains limited in this literature. The importance of embedding context within visualizations is particularly evident in the evolving landscape of science communication, which increasingly unfolds online and via social media \citep{rachaeldavies2021landscape}. This shift is reflected in the growing adoption of a `mobile-first' approach to visualization design in journalism \citep{engebretsen2018data}. In such environments, charts are easily detached from their original context as they circulate on social media and are used as argumentative instruments in public discourse \citep{schneider2012climate,pandey2014persuasive}, as observed during the COVID-19 pandemic \citep{podkul2020coronavirus}. Misinformation research further shows how misleading charts can foster inaccurate interpretations while maintaining an illusion of data-driven insight \citep{lisnic2023misleading}, a risk that increases when contextual information is reduced. Scientific American staff described practices to ensure that visualizations remain interpretable when detached from an article, often including self-contained explanations, annotations, and informative titles.

The comparison between articulated and intended messages showed that producers tend to formulate higher-level narratives, whereas we, as readers, often refer to more concrete details, indicating different levels of abstraction. This resonates with findings that expert-formulated messages tend to be more abstract than those articulated by laypeople \citep{schuster2024being}, pointing to differences in sensemaking processes. Overall, our articulated messages reflected divergence in emphasis, framing, or level of detail rather than systematic misunderstanding, emphasizing that visualizations need not convey a single correct takeaway. In public science communication, readers are encouraged to explore data and draw their own conclusions. Interpretation is therefore context-dependent, and multiple legitimate takeaways may emerge.

These findings carry implications for evaluation practices. Rather than relying solely on accuracy-based readability metrics, our results suggest that evaluation should also consider message-based or insight-based approaches aligned with producers' communicative goals. At the same time, insights are inherently more difficult to measure than traditional understandability metrics, particularly because divergence from the intended takeaway does not necessarily indicate failure. This highlights the need for future research on integrating message-based evaluation into everyday design and evaluation practices. Tools could support this process by enabling collaborative message planning and alignment workflows that allow producers to articulate explicit takeaways early in production and assess whether resulting visualizations support these goals. While user testing remains the preferred approach, journalistic practice often lacks the resources to implement it \citep{schuster2026practitioners}. In such contexts, AI-supported tools could help bridge this gap by generating candidate takeaway messages from visualization drafts, which can then be compared with the intended messages.

\subsection{Studying messages as a methodological contribution} 
\label{subsec:discussion-methodcontribution}

This work also makes a methodological contribution. We used the written articulation of messages to surface traces and signs of sensemaking processes. By externalizing interpretation in written form, articulated messages provide a comparable analytic unit for examining how meaning is constructed across readers, charts, and communicative contexts. We argue that these processes are more complex than what can be revealed through common methods used to evaluate visualizations, such as performance in perceptual operations or knowledge-type tasks \citep{franconeri2021science}. Studying messages, therefore, enables analysis of communicative meaning in more realistic settings rather than isolated task performance. In line with work in education and cognitive science \citep{hidi1986producing,bloom1956taxonomy}, our analysis illustrates that articulating messages involves complex thinking (i.e., in the case of writing more general messages (Section~\ref{subsec:findings-reader})), and that different levels of cognitive activities are interwoven throughout visual data sensemaking.

In addition to writing messages, we also wrote field notes. Writing field notes facilitated deeper cognitive engagement, i.e., analyzing, evaluating, and synthesizing \citep{bloom1956taxonomy}. Methodologically, field notes enabled us to capture contextual and interpretive layers that extend beyond the chart itself. Our field notes also revealed connections between the visualizations, our personal experiences, and the sociopolitical contexts of the charts. This level of content is vital in creating descriptions of charts to facilitate engagement (Section~\ref{subsec:discussion-chart}), to represent a higher level of semantic content \citep{lundgard2022accessible}, and to surface `placing activities' as signs of sensemaking with data \citep{koesten2021talking}. Combining field notes with message articulation allowed this content to emerge in our analysis.

Focusing on our sensemaking processes also allows us to incorporate reflexivity into our research process \citep{berger2015now} and reflect on our positionality. As educated, science-interested people, we align with the envisioned audience of Scientific American, as mentioned in the interviews. We realize, however, that we are shaped by our own experiences, including our experience of being academic researchers in the broad areas of data visualization and human-computer interaction. While some may see this experience as equipping us to understand data visualizations readily, our findings demonstrate that we often struggled to understand visualizations and derive messages. Rather than starting with a larger pool of `lay participants' in our study, we began by questioning what we could understand as people with visualization experience. While this is also a limitation of our work, our findings show that even `experts' have difficulty understanding visualizations, particularly those without additional textual elements that add context. This further underscores the blurry line between lay and expert audiences in science communication \citep{koizumi2021deficit,hilgartner1990dominant}. In future studies, it would be interesting to apply this methodology to examine messages with lay participants from diverse backgrounds. 

Our interviews and message comparison provided a final methodological perspective. 45\% of messages were fully aligned, while 38\% were partially aligned and 17\% diverged substantially from the intended messages of the chart producers. This comparison demonstrates how articulated messages can serve as a bridge between producer intent and reader interpretation, enabling systematic examination of communicative alignment. Our findings underline the importance of testing the success of intended messages with audiences. A more structured evaluation approach, informed by a user-centered design process, could be a further catalyst for more inclusive visualization design in general, potentially encouraging communication between visualization producers and readers to benefit both. Such approaches are particularly valuable in public-facing communication contexts, where effective data communication depends on how messages are interpreted by diverse audiences.

\subsection{Limitations}
\label{subsec:discussion-limitations}
As with all research, this exploratory work has limitations, particularly along the following three dimensions:

\textbf{Scope of the corpus and topics studied.} The selection of Scientific American as a case study reflects a traditional US science communication outlet with a specific editorial context and audience. Our sample of charts is not representative and includes fewer pandemic-related visualizations than climate change visualizations. Both climate change and COVID-19 are topics people feel strongly about, and we expected high rates of bias and preconceptions, which may have influenced our observations of subjective elements and engagement. In addition, the prominence of temporal and geospatial representations in our corpus likely reflects the nature of these topics and may differ in other domains. Future work is needed to investigate how topic characteristics shape message types and interpretation. Findings may also vary in other communication contexts, such as social media graphics, dashboards, or policy communication.

\textbf{Researcher perspective and audience representation.} As discussed in Section~\ref{sec:methodology}, we played a dual role when creating the messages, acting both as researchers in a reflective, analytic mode and as readers of the visualizations. Although we reflected extensively on these roles throughout the research process, this perspective entails two aspects to consider: (a) our visualization expertise and (b) our engagement as researchers. While we fall within the Scientific American target audience of educated, science-interested readers (as confirmed in the interviews), we likely have a higher level of data visualization expertise. Because message articulation required sustained attention and reflection, our engagement with the visualizations, in terms of the time and effort spent understanding them, likely differed from everyday encounters or study settings, where readers often interpret charts quickly and under time constraints or with limited attention. We acknowledge that both aspects may have influenced the messages. In particular, the proportion of misinterpretations may be higher in more typical reading situations, where readers spend less time engaging with visualizations and may not share the same visualization expertise or analytic perspective.

\textbf{Methodological constraints of message articulation.} Lastly, we acknowledge that the process of message articulation captures expressed interpretation and represents only an approximation of tacit insight. The task of verbalizing a takeaway message encourages viewers to summarize and potentially abstract information in ways that may differ from everyday chart encounters. Our findings, therefore, reflect reflective sensemaking through deliberate, effortful engagement. While we argue that takeaway messages provide a valuable lens on insight beyond singular accuracy metrics, they should be interpreted in light of these methodological constraints.

\section{Implications for designing public-facing visualizations}
\label{sec:recommendations}
Taken together, the findings illustrate how visualization messages emerge through interactions among design choices, textual framing, and audience sensemaking. Drawing on patterns identified across the findings, we derive a set of recommendations to support intentional message design, contextual clarity, and audience-centered communication. Rather than universal rules, these recommendations are intended to inform considerations for data visualization producers and science communicators in chart design, many of which are already in practice at Scientific American. Relevant findings sections are indicated in brackets after each recommendation.

\begin{enumerate}
    \item \textbf{\textit{Start with the message}: Articulate the intended message prior to chart design.}
    Defining the communicative goal early helps guide design decisions and ensures coherence between visualizations and their narrative context. Because readers derive messages in context, alignment among the chart, its textual elements, and the surrounding article helps readers construct the intended meaning. Early coordination among editors, writers, and designers supports this consistency. When working with unfamiliar data, exploratory visualization can help producers identify patterns before committing to a message. \textit{(\ref{subsec:findings-producer-planning})}

    \item \textbf{\textit{Be intentional about design}: Be aware of how chart design choices shape interpretation.}
    Design choices afford certain ways of seeing and influence the messages that visualization readers understand. Decisions about aggregation, level of detail, visual emphasis, omission, and textual framing shape both what is perceived and how it is interpreted. Titles, captions, and visual emphasis can position data rhetorically and shift tone, while omissions or selective baselines may unintentionally mislead. Striking this balance is critical for clear and responsible communication and for ensuring that visualizations reflect what the data support. \textit{(\ref{subsec:findings-reader}, \ref{subsec:findings-chart-messagechart}, \ref{subsec:findings-chart-messagetext}, \ref{subsec:findings-producer-planning}, \ref{subsec:findings-producer-comparison})}
    
    \item \textbf{\textit{Guide the eye:} Make the intended takeaway explicit and reduce visual complexity.} 
    Visualizations should clearly signal what viewers are meant to take away. Visual emphasis, layout, and concise textual cues should direct attention toward the central message and prevent secondary elements from becoming unintended focal points. Dense layouts, excessive layering, and multiple simultaneous dimensions increase cognitive effort and hinder synthesis. Simplifying structure, separating comparisons, and balancing detail with clarity support pattern recognition. For composite charts and small multiples, summary statements or small headings that tie components together help readers synthesize across parts and reduce the risk of additive, component-by-component interpretation at the expense of the overall message. \textit{(\ref{subsec:findings-reader}, \ref{subsec:findings-chart-characterization}, \ref{subsec:findings-chart-messagechart}, \ref{subsec:findings-producer-comparison})}

    \item \textbf{\textit{Text-it}: Rethink the role of text on and around the chart.} 
    Textual elements such as annotations, titles, and captions are key to enabling sensemaking success. They can guide attention, clarify how to read the chart, and explain terminology, units, thresholds, or relationships. Words carry rhetorical weight and can substantially shift the perceived framing of a chart. At the same time, people engage with information differently, and visual comprehension differs cognitively from understanding text. Hence, care is needed to ensure textual elements guide rather than distort interpretation. \textit{(\ref{subsec:findings-chart-messagechart}, \ref{subsec:findings-chart-messagetext}, \ref{subsec:findings-producer-planning})}

    \item \textbf{\textit{Show what differences mean}: Make comparisons visible and interpretable through clear reference points.} 
    Even simple visualizations need clear reference points to support sensemaking. Differences should be easy to perceive and interpretable relative to baselines, thresholds, and contextual frames. This is especially important for temporal and geospatial data. Including explanations of units, terminology, and reference values, alongside key information about measurements or data sources, makes distinctions explicit, and clear reference frames support meaningful comparisons. \textit{(\ref{subsec:findings-reader}, \ref{subsec:findings-chart-messagetext}, \ref{subsec:findings-producer-comparison})}
    
    \item \textbf{\textit{Make it stick:} Make it hard to separate context from a chart.} 
    Especially given the ease of sharing on social media, consider how to ensure the message travels with the visualization, e.g., by making it explicit in captions or titles. Other contextual information that supports sensemaking and transparency, such as citations, data sources, or relevant metadata, should also be embedded within or linked to the chart. \textit{(\ref{subsec:findings-chart-messagetext}, \ref{subsec:findings-producer-planning})}

    \item \textbf{\textit{Test the takeaway}: Evaluate communicative success with audiences.} 
    We recommend involving readers in the design process to support a more user-centered approach to visualization design. Visualizations are interpreted through different levels of knowledge, trust, and beliefs; designers should anticipate varied interpretations and reduce ambiguity to support inclusive understanding. Consider the intended audience and tailor messages accordingly, while also reflecting on how other audiences may interpret them. Techniques such as displaying detail on demand can help balance accessibility for general audiences with depth for expert readers. Evaluating communicative success, not only accuracy or task performance, with representatives of different audiences is integral to assessing visualization effectiveness. \textit{(\ref{subsec:findings-reader}, \ref{subsec:findings-chart-messagetext}, \ref{subsec:findings-producer-comparison})}
    
\end{enumerate}

\section{Conclusion}
This work examined visual data communication in Scientific American to understand how visualization messages are constructed and interpreted in public science communication, focusing on climate change and pandemics. Using a mixed-method, multi-perspective approach, we analyzed communicative meaning across reader, chart, and producer perspectives. By operationalizing takeaway messages as an analytic lens and thematically analyzing articulated messages, we offer a methodological approach and typology for examining visualization understanding beyond accuracy-based metrics. Our findings show that meaning emerges not from visual encodings alone but from the interplay of design choices, textual framing, narrative context, and audience sensemaking. Interviews with Scientific American staff highlighted the central role of message planning and alignment with narrative context and audience needs. These insights underscore the importance of text, intentional message design, contextual clarity, and audience-aware communication. We derive recommendations for effective chart design and position visualization as a contextual, communicative practice, pointing toward message-based evaluation frameworks and tools that support producers in testing communicative intent before publication.

\section{Declaration of generative AI and AI-assisted technologies in the manuscript preparation process}
During the preparation of this work, the authors used Grammarly, GPT 5.3, and Claude Sonnet 4.6 in order to improve language and readability. After using these tools, the authors reviewed and edited the content as needed and take full responsibility for the content of the published article.

\section{CRediT authorship contribution statement}
\textbf{Regina Schuster:} Data curation, Formal analysis, Investigation, Visualization, Writing – original draft, Writing – review \& editing. 
\textbf{Kathleen Gregory:} Conceptualization, Data curation, Formal analysis, Funding acquisition, Investigation, Methodology, Writing – original draft, Writing – review \& editing.
\textbf{Christian Knoll:} Formal analysis, Investigation, Visualization, Writing – review \& editing.
\textbf{Torsten Möller:} Funding acquisition, Resources, Supervision, Writing – review \& editing.
\textbf{Laura Koesten:} Conceptualization, Data curation, Formal analysis, Funding acquisition, Investigation, Methodology, Project administration, Writing – original draft, Writing – review \& editing.

\section{Acknowledgements}
This work has been funded by the Vienna Science and Technology Fund (WWTF)[10.47379/ICT20065]. We would like to thank Scientific American for their collaboration and participation in this study.




\bibliographystyle{elsarticle-harv}
\bibliography{bibliography}
\end{document}